\documentclass[epsf]{aastex}
\usepackage{apjfonts}
\usepackage[onecolumn]{emulateapj5}
\usepackage{graphicx} 
\usepackage{color} 
\singlespace
 
\newcommand{\be}{  \begin{eqnarray} }
\newcommand{\ee}{  \end{eqnarray} }

\def\spose#1{\hbox to 0pt{#1\hss}}
\def\lta{\mathrel{\spose{\lower 3pt\hbox{$\mathchar"218$}}
     \raise 2.0pt\hbox{$\mathchar"13C$}}}
\def\gta{\mathrel{\spose{\lower 3pt\hbox{$\mathchar"218$}}
     \raise 2.0pt\hbox{$\mathchar"13E$}}}
\font\syvec=cmbsy10                        %for boldface nabla
\font\gkvec=cmmib10                         %for boldface lowercase Greek
\def\bnabla{\hbox{{\syvec\char114}}}       %bold face nabla
\def\bxi{\hbox{{\gkvec\char24}}}           %bold face xi
\def\bQ{{\bf Q}}        %bold Q
\def\bhat{\hat{\bf b}}        %bold Q

\begin{document}

\shorttitle{Coulomb Bubbles}
\title{Coulomb Bubbles:  Over-stable driving of Magnetoacoustic 
Waves due to the 
Rapid and Anisotropic Diffusion of Energy}
\author{Aristotle Socrates,\altaffilmark{1,2} Ian J. Parrish
\altaffilmark{1} and 
James M. Stone \altaffilmark{1}}
\altaffiltext{1}{Department of Astrophysical Sciences, Princeton 
University, Peyton Hall-Ivy Lane, Princeton, NJ 08544; 
socrates@astro.princeton.edu}
\altaffiltext{2}{Hubble Fellow}

\begin{abstract}

We perform a linear magnetohydrodynamic perturbation analysis for a
stratified magnetized envelope where the diffusion of heat is mediated
by charged particles that are confined to flow along magnetic field
lines.  We identify an instability, the ``coulomb bubble
instability,'' which may be thought of as standard magnetosonic 
fast and slow waves, driven by the rapid diffusion of heat along the
direction of the magnetic field.  We calculate the growth rate and
stability criteria for the coulomb bubble instability for various
choices of equilibrium conditions.

The coulomb bubble instability is intimately related to the 
photon bubble instability.  The bulk thermodynamic
properties of both instability mechanisms are quite similar in that
they require the timescale for heat to diffuse across a wavelength
to be shorter than the corresponding wave-crossing time.  Furthermore,
over-stability occurs only as long as the driving resulting
from the presence of the background heat flux can overcome diffusive
Silk damping.  However, the geometric and therefore mechanical
properties of the coulomb bubble instability is the
complete {\it mirror opposite} of the photon bubble instability.

The coulomb bubble instability is most strongly driven for weakly
magnetized atmospheres that are strongly convectively stable.  We
briefly discuss a possible application of astrophysical interest:
diffusion of interstellar cosmic rays in the hot $T\sim 10^6$ K
Galactic corona.  We show that for commonly accepted values of the
cosmic ray and gas pressure as well as its overall characteristic
dimensions, the Galactic corona is in a marginal state of stability
with respect to a cosmic ray coulomb bubble instability.  The
implication being that a cosmic ray coulomb bubble instability plays a
role regulating both the pressure and transport properties of
interstellar cosmic rays, while serving as a source of acoustic 
power above the galactic disk.

\end{abstract}


\section{Introduction}
   
The idea that the diffusive flow of energy up through an atmosphere
carries with it, the potential to de-stabilize acoustic motion, has
been considered for quite some time (Baker \& Kippenhahn 1962).  The
most familiar situation is where the flow of energy is transmitted by
a radiative heat flux and the instability mechanism operates due to
changes in the opacity that result from nearly adiabatic changes in
temperature and density, along the wave.  The instability mechanism,
known as the $\kappa$-mechanism, is responsible for the strong
observed pulsations in RR-Lyrae and Cepheid variable stars.

In the case where the instability is strong such that the 
pulsation amplitude is large, the overall structure and evolution 
of the equilibrium flow may be significantly altered.  If
the saturation amplitude is small, information on the overall 
structure of the system may be extracted.  

More recently, the inclusion of magnetic fields into perturbation
analyses of optically thick, radiating, stratified flows has produced
some interesting results.  In particular, accretion flows
onto black holes and neutron stars are unstable to the photon bubble
instability (Arons 1992; Gammie 1998; Blaes \& Socrates 2001).
Furthermore, the photon bubble instability, which can be thought of as
a standard magnetosonic wave that is driven over-stable by the
presence of a stratified radiation field, may operate in equilibria
that are weakly magnetized and/or highly sub-Eddington (Blaes \&
Socrates 2003).

Balbus (2001) showed that incompressible Brunt-Vaisalla
oscillations are unstable when horizontally-separated fluid elements
are thermally-connected by coulomb conduction along
field lines of relatively little dynamical importance.
This ``magneto-thermal instability'' operates as long
as the diffusion of heat along the field lines is 
rapid in comparison to the Brunt-Vaisalla frequency.  For 
sufficiently weak fields, instability occurs for
atmospheres with an outward decreasing temperature profile,
rather than an outward decreasing entropy profile -- as in the 
case of convective stability (see e.g. Parrish \& Stone 2005; 
Chandran \& Dennis 2006).

In this work we explore the possibility that magnetoacoustic 
waves may be secularly driven when the envelope's flux of heat 
is mediated by anisotropic thermal diffusion that occurs solely 
along magnetic field lines.  The stability of magnetized 
neutron stars, cosmic ray diffusion in the interstellar medium
and the thermal structure of cluster-scale cooling flows are some 
examples of where the following instability analysis may be be 
relevant.     

The plan of this work is as follows.  In \S\ref{s: basic} we express
our basic assumptions, write down the necessary conservation laws and
then specify the parameters of the simple equilibria, which we then
perturb.  Furthermore, in \S\ref{ss: regimes} we place our analysis
within the context of previous work on linear MHD theory where the
effects of rapid heat diffusion was taken into account. In \S\ref{s:
thermodynamics} our linear analysis begins, with an emphasis on the
thermodynamics of the coulomb bubble instability.  In \S\ref{s:
mechanics}, our linear analysis continues with an emphasis on the
mechanics of the coulomb bubble instability.  In \S\ref{ss: essence},
we present a physical description of the coulomb bubble driving
mechanism in juxtaposition with the photon bubble instability.  In
\S\ref{s: examples}, we discuss the possibility that the Coulomb
Bubble instability might play a role in determining the nature of
interstellar cosmic ray diffusion.  In \S\ref{s: summary} we summarize
our results and discuss possibilities for further research.

\section{Assumptions, Fundamental Equations and Perturbations}
\label{s: basic}

\subsection{Conservation Laws and the Background}

We assume that the basic equations of ideal MHD apply.  Also, 
the only source term in the first law of thermodynamics is
provided by a diffusive heat flux.  The conservation laws given 
below are the same as those found in Balbus (2001); they are
\be
\frac{\partial\rho}{\partial t}+\bnabla\cdot\left(\rho{\bf v}\right)
=0
\label{e: continuity}
\ee
\be
\rho\left(\frac{\partial {\bf v}}{\partial t} +{\bf v}\cdot\bnabla
{\bf v}\right)=-\bnabla P+\rho\,{\bf g}+\frac{1}{4\pi}\left(\bnabla 
\times {\bf B}\right)\times{\bf B},
\label{e: momentum}
\ee
\be
\rho\,T\left(\frac{\partial s}{\partial t}+{\bf v}\cdot\bnabla s\right)
=-\bnabla\cdot{\bf Q}
\label{e: firstlaw},
\ee
\be
{\bf Q}=-\chi{\hat{\bf b}}{\hat{\bf b}}\cdot\bnabla T,
\label{e: heat flux}
\ee
\be
\frac{\partial{\bf B}}{\partial t}=\bnabla\times\left({\bf v}\times
{\bf B}\right),
\label{e: fluxfreeze}
\ee
and
\be
\bnabla\cdot{\bf B}=0.
\label{e: gauss}
\ee
Eqs. (\ref{e: continuity})-(\ref{e: firstlaw}) enforce conservation 
of mass, momentum, and energy, respectively 
while eq. (\ref{e: fluxfreeze}) enforces
magnetic flux-freezing.  Furthermore, 
eq. (\ref{e: heat flux}) constrains
that the flow of heat follows a diffusion law 
along magnetic field lines and 
eq. (\ref{e: gauss}) enforces ${\bf B}$ to be purely 
solenoidal.  Definitions of 
the various symbols in the expressions above, as well as others
used throughout, are listed in 
Table \ref{tbl-0}.

\begin{deluxetable}{cc}
%\rotate
\tabletypesize{\scriptsize}
\tablecaption{Definitions of highly-used symbols
%\tablenotemark{a}. 
\label{tbl-0}}
\tablewidth{0pt}
\tablehead{\colhead{Symbol} &\colhead{Quantity}}
\startdata
$\rho$ & density 
\\
${\bf v}$ & velocity
\\
$P$ & pressure
\\
$T$ & temperature
\\
$s$ & entropy per unit mass
\\
${\bf B}$ & magnetic flux density
\\
$\hat{\bf b}$ & magnetic field unit vector
\\
${\bf Q}$ & heat flux
\\
$\chi$ & thermal conductivity along field lines
\\
${\bf g}$ & gravitational acceleration
\\
$\omega$ & wave frequency
\\
${\bf k}$ & wave vector
\\
$c_i$ & isothermal sound speed
\\
${\bf v_A}$ & Alfv\'en velocity 
\\
$\kappa$ & opacity
 \\
$ \omega_{\rm C,diff}  $ & $\omega+ i\frac{\chi\left({\bf k}\cdot
\hat{\bf b}\right)^2}{n\,T\,\left(\partial s/\partial T\right)_{\rho}}$
 \\
${\tilde\omega}^2$ & $\omega^2-\left({\bf k}\cdot{\bf v_A}\right)^2$
\\
$\xi$ & Lagrangian displacement
\\
$\Delta=\delta+\xi\cdot\nabla$ & Lagrangian variation
\\
$A_0$ & asymptotic (in $k$) growthrate
\\
$\delta{\bf Q}_{\perp}$ & Eulerian heat flux perturbation $\perp {\bf B}$
\enddata 
%\tablenotetext{a}{}
\end{deluxetable}

The form of the heat flux given by eq. (\ref{e: heat flux}) is valid
as long as two conditions are satisfied.  First, the Larmor radius
must be small in comparison to the mean free path of the particles
endowed with relatively large amounts thermal mobility and 
therefore, ${\bf Q}\parallel\hat{\bf b}$.  Second, the mean free
path along ${\bf B}$ must be smaller than the characteristic scales
of the problem e.g., the temperature scale height.  Later on, when we
consider perturbations, the mean free path along the ${\bf B}$ must be
smaller than the wavelength of the perturbation in question.   

We assume that the background is in hydrostatic balance and the 
equilibrium magnetic field ${\bf B}$ is a constant.  We have
\be
-\frac{1}{\rho}\bnabla P={\bf g},
\label{e: hydrostat}
\ee
where changes in the gas pressure $P$ are related to changes in density 
and temperature via an equation of state
\be
P=\frac{\rho\, k_BT}{\mu m_p}.  
\label{e: EOS}
\ee
Throughout, we assume that the mean molecular weight $\mu$ is a constant
so that changes in composition are ignored.  We only consider a 
background that may be characterized as a ``stellar envelope.''  That is, 
both the surface gravity and thermal energy flux throughout the medium is
set to be a constant.  In order to resemble a stellar envelope,
the condition for radiative equilibrium must read
\be
\bnabla\cdot{\bf Q}=0
\label{e: radeq}
\ee
and in addition, the mass of the atmosphere must be insignificant in 
comparison to the source of the gravitational field.  

\subsection{Nature of Perturbations in Different 
Regimes}
\label{ss: regimes}

Our background is simple and closely resembles a basic stellar
atmosphere, threaded by an equilibrium magnetic field.
The only major difference is that energy does not flow upwards via
radiative diffusion, directly against the pull of gravity.  Instead,
the transfer of energy is mediated by charged particles of relatively
high thermal mobility that drift along the equilibrium field.

Now we consider dynamical fluctuations.  To keep things simple, 
we examine local WKB perturbations whose time and spatial 
dependence are $\propto e^{i\left({\bf k}\cdot{\bf x}- 
\omega t\right)}$.  Furthermore, we only consider wavevectors
that are two dimensional i.e.,  ${\bf k}=(k_x,k_z)$, where 
$\hat{\bf z}$ lies in the vertical direction.           
    
\subsubsection{Adiabatic Fluctuations: Alfv\`en, Gravity and 
Magnetosonic Waves}

In the limit of slow thermal diffusion, linear perturbations
to the first law of thermodynamics, eq. (\ref{e: firstlaw}), 
take on a simple form
\be
-i\omega\rho T\left( \delta s+\bxi\cdot\bnabla s \right)=
-i\omega\rho T\Delta s\simeq 0.
\ee
The Eulerian component of $\Delta s$ leads to the fluid's 
acoustic response for short-wavelength perturbations, while
the Lagrangian component $\propto\bxi\cdot\bnabla s$ 
leads to the incompressible gravity waves.  

Including magnetic forces, compressible 
(${\bf k}\cdot\delta{\bf v}\neq 0$) short-wavelength perturbations
obey the magnetosonic dispersion relation given by
\be
\omega^2=k^2\,c^2_s\frac{{\tilde\omega}^2}{\omega^2}+k^2v^2_A,
\label{e: magnetosonic}
\ee
where $c_s$ is the adiabatic sound speed and all other symbols
are defined in Table \ref{tbl-0}.  The eigenvectors 
corresponding to the roots of the dispersion relation above
are referred to as the fast and slow magnetosonic waves.

Incompressible perturbations, with ${\bf k}\cdot\delta{\bf v}=0$,
satisfy the dispersion relation
\be
{\tilde\omega}^2-\frac{k^2_x}{k^2}N^2_{BV}=0
\ee
where $N^2_{BV}$ is the Brunt-Vaisalla frequency.  Basically, the restoring 
force is provided by a combination of Alfv\'enic tension 
and buoyancy.    

%CHECK THE INCOMPRESSIBLE CASE

\subsubsection{Highly Non-Adiabatic Incompressible Fluctuations: 
Balbus' 2001 Analysis}

The fluctuations mentioned above are quite standard and do not
therefore, merit further analysis or discussion.  Interesting physical
effects, such as damping and instability, occur when fluctuations with
respect to the flow of matter and the flow of heat separate from one
another i.e., when the perturbations become non-adiabatic.
Somewhat counter-intuitively, non-adiabatic effects occur when the
time scale for heat flow over a wavelength is shorter than the
oscillation period of the perturbation in question.  That is, the flow
becomes non-adiabatic once the agent of heat transfer can either inject
into, or remove energy from, the flow before the fluid has time to
respond.

Balbus (2001) studied incompressible waves (${\bf k}\cdot\delta{\bf
v}=0$ and $c_s\rightarrow\infty$) in the limit where the diffusion
time along the magnetic field is short in comparison to the wave
crossing time.  Another condition for Balbus' (2001) instability is
that the Alfv\'en time must be shorter than the wave crossing time so
that Alfv\'enic tension cannot suppress the unstable growth of the
perturbations.  He found that Brunt-Vaisalla oscillations (or {\it
g}-modes) are driven unstable if horizontally-separated fluid elements
can quickly transfer heat from regions of relatively high to low
temperatures, once they are perturbed in opposite directions along the
background temperature gradient.  The role of the magnetic field in
this ``magneto-thermal instability'' is to serve as conduit of thermal
energy between horizontally-separated fluid elements.

\subsubsection{Highly Non-Adiabatic Compressible Fluctuations: 
This work}

We now consider the manner in which compressible 
(slow and fast) magnetosonic waves are affected by the rapid flow of 
heat along magnetic field lines.  In a sense, the instability and 
damping mechanisms covered in the next few sections can be viewed
as a natural extension of either Balbus' (2001) or Blaes \& 
Socrates' (2003) analysis of magnetoacoustic waves driven by the 
rapid diffusion of heat along the radiation pressure gradient.  

\section{Total Pressure Perturbation in the Limit of Rapid  
Heat Conduction}\label{s: thermodynamics}

The rapid diffusion of heat alters standard
magnetoacoustic motion through the pressure perturbation, which we 
write as
\be
\delta P=\left(\frac{\partial P}{\partial\rho}\right)_{T}\delta\rho
+\left(\frac{\partial P}{\partial\,T}\right)_{\rho}\delta T.
\label{pressure}
\ee  
That is, we assume that the fluid's pressure is a function of 
of $\rho$ and $T$ only such that $P=P(\rho,T)$.  Furthermore,
$\left({\partial P}/{\partial\rho}\right)_{T}=c^2_i$
where $c_i$ is the isothermal sound speed.  For short-wavelength
magnetosonic motion, $\delta\rho$ is constrained by the linearized 
expression for conservation of mass, while $\delta T$ is constrained by 
the linearized first law of thermodynamics, which reads
\be
-i\,\omega\,n\,T\left(\delta s+\bxi\cdot\bnabla s\right)
=-i{\bf k}\cdot\delta{\bf Q},
\label{1stlaw}
\ee 
where $s$ and $\delta{\bf Q}$ is the specific entropy and the 
perturbed coulombic heat flux, respectively.  Later on, we 
show that the manner
in which $\delta{\bf Q}$ responds to changes in magnetofluid
variables ultimately
determines the nature of the various driving and damping mechanisms.  
Below, we carefully discuss the form of $\delta{\bf Q}$.

\subsection{The Heat Flux Perturbation}
\label{ss: heat flux}

The equilibrium coulombic heat flux is given by 
\be
\bQ=-\chi\bhat\bhat\cdot\bnabla T=-{\bf X}\cdot\bnabla T.
\ee 
Here, $\chi$ is the conductivity along field lines which lie 
in the direction ${\bf B}/B=\bhat$ and ${\bf X}=
{\bf X}(\chi,\bhat)$ is the thermal conductivity tensor. Upon 
perturbation, we have
\be
\delta{\bf Q}=-\delta{\bf X}\cdot\bnabla T-{\bf X}\cdot\bnabla\delta T 
=-\chi\left[\frac{\delta\chi}{\chi}\bhat\left(\bhat\cdot\bnabla T\right) 
+\delta\hat{\bf b}\left(\hat{\bf b}\cdot\bnabla T\right)
+\hat{\bf b}\left(\delta\hat{\bf b}\cdot\bnabla T\right) +
i\,\hat{\bf b}\left({\bf k}\cdot\hat{\bf b}\right)\delta T\right]   
\label{deltaQ1}
\ee
where the coulombic conductivity $\chi$ is not taken to be a constant.
\footnote{Note that the thermal coulomb conductivity 
$\chi$ was set to a constant in Balbus' (2001) analysis.}
We choose to parameterize the conductivity $\chi$ in the following 
way,
\be
\chi=\chi_0T^n\rho^m\kappa^p
\ee
where $\chi_0$ is a constant and $\kappa$ may be thought of as 
an ``opacity'' or in other words, a cross section per unit mass.  Comparison 
with thermally diffusive damping and driving mechanisms for backgrounds 
in which radiation mediates the flow of energy is facilitated if we choose
$n=3$, $m=-1$, and $p=-1$ as in the case of radiative diffusion.  In a
way, we absorb the micro-physical differences (such as cross section)
between radiative and coulombic diffusion into an opacity law
-- or a mean free path --  for the particles that possess relatively 
large amounts of thermal mobility.

From the above considerations, the expression for the conductivity 
perturbation is given by
\be
\frac{\delta\chi}{\chi}=-\frac{\delta\rho}{\rho}+3\frac{\delta T}{T}
-\frac{\delta\kappa}{\kappa}.
\ee
For now, we somewhat artificially set $\delta\kappa=0$, which momentarily 
prevents us from studying a potential $\kappa-$mechanism that results 
from coulomb diffusion.  With this, the expression for the heat flux 
perturbation becomes
\be
\delta{\bf Q}=-\chi\left[-\frac{\delta\rho}{\rho}\,\bhat
\left(\bhat\cdot\bnabla T\right) 
+\delta\hat{\bf b}\left(\hat{\bf b}\cdot\bnabla T\right)
+\hat{\bf b}\left(\delta\hat{\bf b}\cdot\bnabla T\right) +
i\,\hat{\bf b}\left({\bf k}\cdot\hat{\bf b}\right)\delta T\right]
\label{deltaQ2}.   
\ee
Note that we have dropped the term in $\delta\chi\propto 3\,\delta T$
since it contributes at a lower order in $(kH)^{-1}$ relative to the
last term in the above expression, within the context of our WKB
approximation.  We do this with hindsight earned from Blaes \&
Socrates' (2003) analysis.  The rapid diffusion (high-$k$ limit) of
heat along temperature gradients necessarily implies that the
temperature perturbation is relatively small by a factor of $1/k$ in
comparison perturbations of other magnetofluid quantities such as
$\delta\rho$, $\delta\bhat$ and $\delta {\bf v}$.

\subsection{The Temperature Perturbation}
\label{ss: deltaT}

By expanding eq. (\ref{1stlaw}), the first law of thermodynamics,
with the help of 
the expression for the heat flux perturbation $\delta\bQ$ given by 
eq. (\ref{deltaQ2}), we have
\be
-\omega\,n\,T\left[\left(\frac{\partial s}{\partial\rho}\right)_{T}
\delta\rho+ \left(\frac{\partial s}{\partial\,T}\right)_{\rho}
\delta\,T+\bxi\cdot\bnabla s \right]=\chi\left[
-\frac{\delta\rho}{\rho}\left({\bf k}\cdot\bhat\right)
\left(\bhat\cdot\bnabla T\right)+
\left({\bf k}\cdot\delta\hat{\bf b}\right)
\left(\hat{\bf b}\cdot\bnabla T\right)
+\left({\bf k}\cdot\hat{\bf b}\right)
\left(\delta\hat{\bf b}\cdot\bnabla T\right) +
i\left({\bf k}\cdot\hat{\bf b}\right)^2\delta T\right] 
\ee  
It is useful to define a characteristic diffusion frequency,
$\omega_{\rm C,diff}$, which quantifies the rate at which 
heat diffuses over a wavelength,
\be
\omega_{\rm C,diff}\equiv\omega + i\frac{\chi\left({\bf k}\cdot
\hat{\bf b}\right)^2}{n\,T\,\left(\partial s/\partial T\right)_{\rho}}.
\ee
With this, we temperature perturbation becomes
\be
\delta T =-\frac{\omega}{\omega_{\rm C,diff}}\frac{\left(\partial s/
\partial\rho\right)_T}{\left(\partial s/\partial T\right)_{\rho}}
\delta\rho-\frac{\omega}{\omega_{\rm C,diff}}\frac{\bxi\cdot\bnabla
s}{\left(\partial s/\partial T\right)_{\rho}}-\frac{\chi\left[
-\frac{\delta\rho}{\rho}\left({\bf k}\cdot\bhat\right)\left(\bhat\cdot
\bnabla T\right)+
\left({\bf k}\cdot\delta\hat{\bf b}\right)
\left(\hat{\bf b}\cdot\bnabla T\right) +  \left({\bf k}
\cdot\hat{\bf b}\right)\left(\delta\hat{\bf b}\cdot\bnabla T\right)\right]}{
\omega_{\rm C,diff}\,n\,T\left(\partial s/\partial T\right)_{\rho}}.
\ee
By knowing beforehand that the magnetosonic waves of interest 
can be treated to lowest order
as standard fast and slow magnetoacoustic waves tells us that
$\omega\simeq v_{\rm ph}\,k$, where $v_{\rm ph}$ is the phase
velocity of the given wave.  Therefore, in the rapidly diffusing 
limit,     
\be
|\omega_{\rm C,diff}|\rightarrow \frac{\chi\left({\bf k}\cdot
\hat{\bf b}\right)^2}{n\,T\left(\partial s/\partial T\right)_{\rho}}
\gg\omega.
\ee
Now, the temperature perturbation reduces to 
\be
\delta T & \simeq &\frac{i}{\left({\bf k}\cdot\hat{\bf b}\right)^2}
\left[\frac{n\,\rho\,T\omega}{\chi}
\left(\frac{\partial s}{\partial\rho}\right)_T\frac{\delta\rho}{\rho}
 -\frac{\delta\rho}{\rho}\left({\bf k}\cdot\bhat\right)\left(\bhat\cdot
\bnabla T\right)+
\left({\bf k}\cdot\delta\hat{\bf b}\right)
\left(\hat{\bf b}\cdot\bnabla T\right) +  \left({\bf k}
\cdot\hat{\bf b}\right)\left(\delta\hat{\bf b}\cdot\bnabla T\right)
\right]
\label{deltaT}
\ee
The form of the temperature
perturbation given by eq. (\ref{deltaT}) is remarkably similar to 
the case of rapid radiative diffusion, which potentially
leads to photon bubble-like phenomena  i.e.,
\be
 \delta T & \simeq & 
\frac{i}{\,k^2}
\left[\frac{n\,\rho\,T\omega}{\chi}
\left(\frac{\partial s}{\partial\rho}\right)_T
\frac{\delta\rho}{\rho}-
\frac{\delta\rho}{\rho}\,\left({\bf k}\cdot\bnabla T
\right)\right]
\,\,\,\,\,\,\,\,{\rm RADIATIVE\,\,\,\, DIFFUSION}.
\label{deltaTrad}
\ee  
In both cases, the first term which is $\propto\left(\partial s/
\partial\rho\right)_T$ leads to diffusive thermal damping.  For
the specialized case of radiative diffusion, this effect is known 
as Silk damping (Silk 1968; Weinberg 1971).  The remaining terms 
in eq. (\ref{deltaT}) and  eq. (\ref{deltaTrad}) -- 
the one that contain gradients in temperature -- 
lead to secular over-stable driving.  For photon bubbles driven by 
rapid radiative diffusion, the term $\propto{\bf k}\cdot\bnabla T
\,\delta\rho/\rho$ in the temperature perturbation represents 
``shadowing'' or ``pile up.''  More specifically, immediately 
downstream a density maximum, radiation piles up due to the local 
increase in extinction and since photon number is conserved, a
deficit of radiation, or a shadow, occurs.  Thus, the resulting
radiation pressure differential across the density maxima allows
for the possibility of work being performed on the fluid by 
the radiation field.  In the case of anisotropic coulomb
conductivity, a local change in the orientation of the 
magnetic field either permits or deters the flow of heat across a 
local density maximum.  The orientation of $\delta\hat{\bf b}$
relative to constant density surfaces (that are $\perp{\bf k}$)  
ultimately determine whether or not the change in heat flow drives or
damps the oscillation.         

It follows that the total pressure perturbation is divided into two parts;
one which yields a standard acoustic response $\propto\delta\rho$ 
while the other
component $\propto i\,k^{-1}\delta\rho$ leads to 
conductive driving and damping.  That is,
\be
\delta P=\delta P_{\rm ac}+\delta {\tilde P}
\label{deltaP1}
\ee  
where
\be
\delta P_{\rm ac}=\left(\frac{\partial P}{\partial\rho}\right)_{T}
\,\delta\rho=c^2_i\delta\rho\,\,\,\,\,\,\,&{\rm and}&\,\,\,\,\,\,\,
\delta {\tilde P}=\left(\frac{\partial P}{\partial T}\right)_{\rho}
\,\delta T .
\label{deltaP2}
\ee

\section{Coupling of Magnetosonic Motion to the Background Coulomb
Flux}
\label{s: mechanics}

In what follows, we calculate the growth rates and stability
criteria for the fast and slow coulomb bubble instability. 
We start with a brief overview of standard magnetosonic waves,
where background gradients are ignored.  We derive the basic
properties and reveal the nature of the coulomb bubble
instability by determining the ratio of the work done upon a 
magnetosonic wave by the driving mechanism to the 
wave energy (or wave action), which is equal to the ratio 
of the growth/damping rate to the oscillation frequency of
the wave.  We first arrive at this ``work integral''-like
ratio by examining the product of the Lagrangian pressure 
and density perturbation $\Delta P\Delta\rho$, which 
closely resembles the $-PdV$ work done on the wave.  Then, 
we derive the work done on the wave through the 
quantity $\bxi\cdot\delta{\bf f}$, where $\delta{\bf f}$
is the linear driving force resulting from rapid coulomb 
diffusion.

\subsection{Basic Magnetosonic Waves}

Linearizing the continuity, momentum, and induction equations, as well
as Gauss' Law, yields
\begin{equation} 
-i\omega\delta\rho + i\rho\left({\bf k}\cdot\delta {\bf v}\right) +
\delta {\bf v}\cdot{\bnabla}\rho=0,
\label{dcont}
\end{equation}
\begin{equation}
-i\omega\rho\delta{\bf v}=-i{\bf k}\,\delta P+
{\bf g}\,\delta\rho+{i\over4\pi}({\bf k}\times\delta{\bf B})\times{\bf B},
\label{dgasmom}
\end{equation}
\begin{equation}
-i\omega\delta{\bf B}=i{\bf k}\times(\delta{\bf v}\times{\bf B}),
\label{dfluxfreezing}
\end{equation}
and
\be
{\bf k}\cdot\delta{\bf B}=0,
\label{dgauss}
\ee
respectively.

Blaes and Socrates (2003) provide the mathematical apparatus for
ascertaining the growth rate, stability criteria and basic physics of
magnetosonic waves that are either driven or damped by the rapid
diffusion of energy (see Socrates et al. 2005 and Turner et al. 2005
as well).  Solutions of the entire dispersion relation -- for the case
of radiative diffusion -- show that in the high-{\it k} limit, the
growth rate $A_0$ of the fast and slow wave, resulting from photon
bubble driving, is a constant.  Since the oscillation frequency for
standard magnetosonic waves $\omega\propto k$ the ratio of the
magnitude of the driving force to the magnitude of the magnetosonic
restoring force is $\propto A_0/k\propto 1/k$.  It follows that the
terms responsible for driving and damping in the linearized Euler
equation are those that are $\propto\delta T$, the buoyancy force
$\propto{\bf g}\,\delta\rho$ and the component of the Lagrangian
density perturbation $\propto\bxi\cdot\bnabla{\rm ln}\,\rho$, since
all of these quantities belong to linear forces that are $1/k$ times
smaller than the components responsible for the linear magnetosonic
restoring force.  In order to evaluate the growth rate, we insert the
magnetosonic eigenvector into the $\mathcal{O}\left( k^{-1}\right)$
linear forces, giving us the desired correction to the
oscillation frequency.  The technique outlined above is quite similar
to the work integral approach in classic stellar pulsation theory
(Unno 1989) for obtaining growth rates due to non-adiabatic driving of
nearly adiabatic pulsations or to the method of 
calculating shifts in eigenfrequency in the linear 
perturbation theory of particles in quantum mechanics.

We start by obtaining the form of the basic magnetosonic 
eigenvectors and their respective eigenfrequencies.  
In the high$-k$ limit, the dominant terms of the 
Euler equation are
\be
-i\omega\rho\,\delta{\bf v}=-i\,{\bf k}\left[\delta P_{\rm ac}+
\frac{{\bf B}\cdot\delta{\bf B}}{4\pi}\right]+
i\frac{{\bf k}\cdot{\bf B}}{4\pi}\delta{\bf B},
\label{tempeuler}
\ee   
where $\delta P_{\rm ac}\equiv c^2_{\rm i}\delta\rho$ measures
the acoustic pressure response to a perturbation in density $\delta\rho$.   
Furthermore,
\be
\delta\rho=\rho\frac{{\bf k}\cdot\delta{\bf v}}{\omega}\,\,\,\,\,\,\,\,
{\rm and}\,\,\,\,\,\,\,\,
\delta{\bf B}=\frac{{\bf B}}{\omega}\left({\bf k}\cdot\delta{\bf v}
\right) - \frac{{\bf k}\cdot{\bf B}}{\omega}\delta{\bf v},
\label{deltarelations}
\ee
%The fact that the Lorentz force vanishes along the direction of motion
%such that
%\be
%{\bf B}\cdot\delta{\bf v}=c^2_i\frac{{\bf k}\cdot{\bf B}}{\omega}
%\frac{\delta\rho}{\rho}
%\label{Bdotv}
%\ee
%allows us to write
to leading order.  With this, we write $\delta{\bf B}$
in terms of $\delta\rho$ and $\delta{\bf v}$ so that
\be
-i\rho\frac{{\tilde\omega}^2}{\omega}\delta{\bf v}=
-i{\bf k}\left[\left(c^2_i+v^2_A\right)\delta\rho
-c^2_i\frac{\left({\bf k}\cdot{\bf v_A}\right)^2}
{\omega^2}\delta\rho\right]+i\left({\bf k}\cdot{\bf v_A}\right)
{\bf v_A}\,\delta\rho,
\label{tempeuler2}
\ee
where ${\bf v_A}={\bf B}/\sqrt{4\pi\rho}$ is the Alfv\'en velocity
and ${\tilde\omega}^2\equiv\omega^2-({\bf k}\cdot{\bf v_A})^2$.  
In the short-wavelength limit, changes in temperature resulting from 
coulomb conduction force the gas 
via a pressure gradient, which is $\parallel$ to ${\bf k}$ in 
the short wave-length limit.  Therefore, the fluctuations of interest 
must possess some longitudinal, or compressible, 
component which automatically rules out purely 
incompressible shear Alfv\'en wave as a candidate for over-stable 
coulombic driving.  We define a mode polarization  
${\hat{\bf\epsilon}}(\omega,k)$ such that $\delta{\bf v}=
{\hat{\mathbb\epsilon}}(\omega,k)\psi(\omega,k)$ where $\psi$ is some
complex-valued amplitude.  We arbitrarily choose
$\psi=\delta\rho/\rho$ for the magnetosonic eigenvectors
\be
{\hat{\bf\epsilon}}(\omega,k)=\frac{\omega}{{\tilde\omega}^2}\left[
\left(\frac{{\tilde\omega}^2}{\omega^2}\,c^2_i+v^2_A\right){\bf k}-
\left({\bf k}\cdot{\bf v_A}\right){\bf v_A}\right]
\label{polarization}
\ee 
where the two magnetosonic waves must satisfy the dispersion 
relation
\be
\omega^2=k^2\,c^2_i\frac{{\tilde\omega}^2}{\omega^2}+k^2v^2_A.
\label{dispersion}
\ee
In principle, the fast and slow waves posses components both 
$\parallel$ and $\perp$ to ${\bf b}={\bf B}/B={\bf v_A}/v_A$.  

\subsection{The Magnetosonic Wave Equation Subject to
Rapid Coulombic Diffusion:  Asymptotic Growth Rate}

By taking the divergence of eq. (\ref{dgasmom}), we immediately
restrict our analysis to compressible short-wavelength perturbations
i.e., the waves that can be driven unstable by the effects of thermal 
diffusion.  We write
\be
-i\omega{\bf k}\cdot\delta{\bf v}=-ik^2\left[\frac{\delta P}{\rho}
+\frac{{\bf B}\cdot\delta{\bf B}}{4\pi\rho}\right] +\left({\bf k}\cdot
{\bf g}\right)\frac{\delta\rho}{\rho}
\ee
where we have made use of eq. (\ref{dgauss}).  
We eliminate $\delta{\bf B}$ in favor of $\delta{\bf v}$  
with the help of eq. (\ref{dfluxfreezing}), which allows us 
to write
%Again, the fact that
%the Lorentz force vanishes along the direction of motion conveniently
%allows us to eliminate $\delta{\bf B}$ by use of eqs. 
%(\ref{dfluxfreezing})  and (\ref{dgasmom})
\be
-i\omega\left({\bf k}\cdot\delta{\bf v}\right)=-i\,k^2\left[
\frac{\delta P}{\rho}+v^2_A\frac{\left({\bf k}\cdot\delta
{\bf v}\right)}{\omega} -\frac{\left({\bf k}\cdot
{\bf v_A}\right)^2}{\omega^2}\frac{\delta P}{\rho}
-i\frac{\left({\bf g}\cdot{\bf v_A}\right)\left({\bf k}
\cdot{\bf v_A}\right)}{\omega^2}\frac{\delta\rho}{\rho} \right]
+\left({\bf k}\cdot{\bf g}\right)\frac{\delta\rho}{\rho}.
\ee 
By expanding the pressure perturbation into its acoustic $\delta P_{\rm ac}$ 
and 
driving/damping $\tilde{\delta P}$ contributions and by
utilizing eq. (\ref{dcont}), we separate the standard magnetosonic
wave equation from terms that may lead to driving and damping 
\be
\left[\omega^2-k^2\frac{{\tilde\omega}^2}{\omega^2}c^2_i
-k^2v^2_A\right]\frac{\delta\rho}{\rho}\simeq -i\,\omega\delta
{\bf v}\cdot\bnabla{\rm ln}\rho+k^2\Biggl\{\frac{{\tilde\omega}^2}
{\omega^2}\frac{\delta{\tilde P}}{\rho}+iv^2_A\frac{\delta
{\bf v}}{\omega}\cdot\bnabla{\rm ln}\rho -i\frac{\left({\bf g}
\cdot{\bf v_A}\right)\left({\bf k}
\cdot{\bf v_A}\right)}{\omega^2}\frac{\delta\rho}{\rho} \Biggr\}
+i\left({\bf k}\cdot{\bf g}\right)\frac{\delta\rho}{\rho}.
\label{sonic1}
\ee
The left hand side of eq. (\ref{sonic1}) is the magnetosonic wave
equation for a uniform background.  The terms on the right hand side
of eq. (\ref{sonic1}) produce ${\mathcal O}(k^{-1})$ corrections to
the magnetosonic wave eigenvectors and oscillation frequencies as they
are responsible for driving and damping.  In order to evaluate the
frequency correction arising from coulombic diffusion, we let
$\omega\rightarrow\omega +A_0$, where $A_0$ is the damping or driving
rate that is independent of the wavenumber $k$.

The condition for hydrostatic balance, eq. (\ref{e: hydrostat}),
allows us to recast the gravitational acceleration ${\bf g}$ 
in the following useful form
\be
{\bf g}=\frac{1}{\rho}\bnabla P=c^2_i\bnabla{\rm ln}\rho+\frac{1}{\rho}
\left(\frac
{\partial P}{\partial T}\right)_{\rho}\bnabla T.
\label{hydrostatic}
\ee 
Together with the expression for the magnetosonic polarization vector given 
by eq. (\ref{polarization}), we eliminate all of the 
terms $\propto\bnabla{\rm ln}\rho$ in eq. (\ref{sonic1}).  
The perturbative prescription for the eigenvalues
motivates conversion of the altered magnetosonic 
wave equation, given by eq. (\ref{sonic1}), into an expression for the 
damping or driving rate.  After a bit of algebra, we have
\be
\frac{2\,A_0}{\omega}\left[2\omega^2-k^2\left(c^2_i+
v^2_A\right)\right]\frac{\delta\rho}{\rho}\simeq k^2\Biggl\{ 
\frac{{\tilde\omega}^2}{\omega^2}\frac{\delta{\tilde P}}{\rho}
-i\frac{\left({\bf k}\cdot{\bf v_A}\right)}{\omega^2\rho}\,{\bf v_A}
\cdot\left[\left(\frac
{\partial P}{\partial T}\right)_{\rho}
\bnabla T \right]\frac{\delta\rho}{\rho}
\Biggr\}
+i\frac{{\bf k}}{\rho}\cdot\left[ \left(\frac
{\partial P}{\partial T}\right)_{\rho}
\bnabla T \right]\frac{\delta\rho}{\rho}.
\label{sonic2}
\ee  
By noting the form of the magnetosonic polarization vector given by
eq. (\ref{polarization}), we may write
\be
\frac{2\,A_0}{\omega}\left[2\omega^2-k^2\left(c^2_i+
v^2_A\right)\right]\frac{\delta\rho}{\rho}\simeq 
\frac{k^2{\tilde\omega}^2}{\omega^2}\frac{1}{\rho}\left(\frac{\partial P}
{\partial T}\right)_{\rho}\Delta T
\ee
where $\Delta T=\delta T+\bxi\cdot\bnabla T$ is the Lagrangian 
temperature perturbation.  The expression for the growth rate given above
may be written into a more a revealing form if we realize that
\footnote{Due to our choice of normalization 
$\left({\bf k}\cdot\hat{\bf\epsilon}\right)=\omega$, a relation that will 
be useful when we derive the growth rate later on.}
\be
\frac{{\bf k}\cdot\hat{\bf\epsilon}}{\hat{\bf\epsilon}\cdot
\hat{\bf\epsilon}}=\frac{{\tilde\omega}^2k^2}{\omega\left[2\omega^2
-k^2\left(v^2_A+c^2_i \right)\right]},
\ee
which allows us to write
\be
\frac{A_0}{\omega}\simeq\frac{1}{2}\frac{\frac{\Delta\rho}{\rho}\left(
\frac{\partial P}{\partial T}\right)_{\rho}\Delta T}{\rho\,\delta{\bf v}
\cdot\delta{\bf v}}
\label{e: simplegrowth}
\ee 
since $\delta{\bf v}=\hat{\bf\epsilon}\delta\rho/\rho$ and $\left(
{\bf k}\cdot\bxi\right)=i\Delta\rho/\rho$.\footnote{Eq. 
(\ref{e: simplegrowth}) applies for radiative 
diffusion as well.  If $\delta T$ is given by eq. (\ref{deltaTrad}),
then eq. (\ref{e: simplegrowth}) yields the photon bubble growth 
rate and stability criteria for the fast and slow magnetosonic 
waves.}  
Eq. (\ref{e: simplegrowth})
has a straightforward interpretation.  The ratio of the growth 
rate to the oscillation frequency, $A_0/\omega$, is equal to the
work done upon a nearly isothermal perturbation by changes in
the flow's coulombic flux of energy, divided by the
energy of that perturbation.   

Before we complete our calculation of the growth rate $A_0$, we 
recast the perturbation of the magnetic field unit vector 
in the following helpful forms
\be
\delta\hat{\bf b}=\hat{\bf b}\times\left(\frac{\delta{\bf B}}{B}\times
\hat{\bf b}\right)=\frac{\delta{\bf B}}{B}\cdot\left({\bf 1}-
\hat{\bf b}\hat{\bf b}\right)=\frac{\delta{\bf B}_{\perp}}{B}=
i\left({\bf k}\cdot\bhat\right)\bxi
\cdot\left({\bf 1}-\bhat\bhat\right) 
=-\frac{\left({\bf k}\cdot\hat{\bf b}\right)}{{\tilde
\omega}^2}\frac{\omega^2}{k^2}\,{\bf k}\cdot\left({\bf 1}-\hat
{\bf b}\hat{\bf b}\right)\frac{\delta\rho}{\rho},
\label{deltabhat}
\ee  
where eqs. (\ref{deltarelations}) 
and (\ref{polarization}) were put to use.  The Lagrangian temperature
perturbation $\Delta T$ may be broken into a part that 
is responsible for diffusive Silk-like damping, which we 
denote as $\Delta T^{\rm damp}$ and a portion $\Delta T^{\rm drive}$ 
which may lead to over-stable driving.  We have 
\be
\Delta T^{\rm damp}=\delta T^{\rm damp}
\simeq i\frac{n\,\rho\,T\omega}{\left({\bf k}\cdot\bhat
\right)^2\chi}
\left(\frac{\partial s}{\partial\rho}\right)_T\frac{\delta\rho}{\rho}
\label{Tdamp}
\ee  
and
\be
\Delta T^{\rm drive}&=&\delta T^{\rm drive}+\bxi\cdot\bnabla T
\simeq\frac{i}{\left({\bf k}\cdot\hat{\bf b}\right)^2}
\left[-\frac{\delta\rho}{\rho}\left({\bf k}\cdot\bhat\right)
\left(\bhat\cdot\bnabla T\right)+
\left({\bf k}\cdot\delta\hat{\bf b}\right)
\left(\hat{\bf b}\cdot\bnabla T\right) +  \left({\bf k}
\cdot\hat{\bf b}\right)\left(\delta\hat{\bf b}\cdot\bnabla T\right)
\right]+\bxi\cdot\bnabla T\nonumber\\
&\simeq& -\frac{1}{\left({\bf k}\cdot\hat{\bf b}\right)^2}\left[
\left({\bf k}\cdot\bxi\right)\left({\bf k}\cdot\bhat\right)\left(
\bhat\cdot\bnabla T\right)+\left({\bf k}\cdot\bhat\right){\bf k}\cdot
\left(\bxi\cdot\left({\bf 1}-\bhat\bhat\right)\right)\left(\bhat\cdot
\bnabla T\right)
+\left({\bf k}\cdot\bhat\right)^2\bxi\cdot\left({\bf 1}-\bhat\bhat\right)
\cdot\bnabla T\right]+\bxi\cdot\bnabla T\nonumber\\
&\simeq&-2\frac{\left(\bhat\cdot\bnabla T\right)}{\left({\bf k}
\cdot\hat{\bf b}\right)^2}\left[
\left({\bf k}\cdot\bxi\right)\left({\bf k}\cdot\bhat\right)
-\left({\bf k}\cdot\bhat\right)^2\left(
\bhat\cdot\bxi\right)\right]=2\frac{{\tilde\omega}^2k^2}{\omega^2} 
\frac{\left(\bhat\cdot\bnabla T\right)}
{\left({\bf k}\cdot\hat{\bf b}\right)^2}
\frac{\bxi\cdot\delta\bhat}{\delta\rho/\rho}=
-2\frac{{\tilde\omega}^2k^2}{\omega^2}
\frac{\bxi\cdot\delta{\bf Q}_{\perp}}{\chi\left({\bf k}\cdot
\bhat\right)^2\delta\rho/\rho}\simeq \Delta T^{\rm drive}
\label{Tdrive}
\ee
where we have made extensive use of eq. (\ref{deltabhat}).  In above 
expression for $\Delta T^{\rm drive}$, the term $\delta{\bf Q}_{\perp}
\equiv\delta{\bf Q}\cdot\left({\bf 1}-\bhat\bhat\right)$ is
the component of the heat flux perturbation $\delta{\bf Q}$
that is perpendicular to the direction of the 
equilibrium field $\bhat$ (and thus ${\bf Q}$).
As eq. (\ref{Tdrive}) implies, the perpendicular heat
flux perturbation $\delta{\bf Q}_{\perp}$ takes the form
\be
\delta {\bf Q}_{\perp}\equiv\left({\bf 1}-\bhat\bhat\right)
\cdot\delta{\bf Q}=-\chi\left({\bf 1} -\bhat\bhat\right)
\cdot\left[\delta\bhat\left(\bhat\cdot\bnabla T\right)\right]
=-\chi\left({\bf 1} -\bhat\bhat\right)
\cdot\left[i\left({\bf k}\cdot\bhat\right)\bxi\cdot
\left({\bf 1} -\bhat\bhat\right)\right]\left(\bhat\cdot
\bnabla T\right)=-\chi\,\delta\bhat\left(\bhat\cdot\bnabla T\right),
\ee
since by construction, $\delta\bhat$ is $\perp$ to $\bhat$.  Thus, the
form of $\Delta T^{\rm drive}$ tells us that work done upon a fluid
element by $\delta{\bf Q}$ is proportional to the projection of the
heat flux perpendicular to the equilibrium field, which is parallel to
the perturbation of the magnetic field direction vector $\delta\bhat$,
upon the fluid displacement $\bxi$.  Ultimately, the perpendicular
heat flux $\delta{\bf Q}_{\perp}$ must correspond to the linear
driving force -- a point that we further consider in the following
section.

Finally, we are in the position to write the growth rate $A_0$ in terms
of background quantities and the wave vector ${\bf k}$ of the 
oscillation
\be
A_0&\simeq&\frac{i\,{\tilde\omega}^2\,k^2}{2\omega\left[2\omega^2-
k^2\left(c^2_i+v^2_A\right)\right]}
\frac{1}{\rho}\left(\frac{\partial P}{\partial T}\right)_{\rho}
\times\Biggl\{\frac{\rho\,n\,T\omega}
{\chi\left({\bf k}\cdot\hat{\bf b}\right)^2}
\left(\frac{\partial s}{\partial\rho}\right)_T 
-2\,\frac{\omega^2}{{\tilde\omega}^2}
\frac{\left(\bhat\cdot\bnabla T\right)}{\left({\bf k}\cdot
\bhat\right)}\left[1-\frac{\left({\bf k}
\cdot\bhat\right)^2}{k^2}\right]
\Biggr\}
\label{growthrate}
\ee
and in terms of the background coulombic heat flux, the growth rate
is given given by
\be
A_0&\simeq&\frac{i\,{\tilde\omega}^2\,k^2}{2\omega\left[2\omega^2-
k^2\left(c^2_i+v^2_A\right)\right]}
\left(\frac{\partial P}{\partial T}\right)_{\rho}
\times\frac{1}{\rho\,\chi\left({\bf k}\cdot\bhat\right)^2}
\Biggl\{{\rho\,n\,T\omega}
\left(\frac{\partial s}{\partial\rho}\right)_T 
+2\,\frac{\omega^2}{{\tilde\omega}^2}\left({\bf k}\cdot
{\bf Q}\right)\left[1-\frac{\left({\bf k}
\cdot\bhat\right)^2}{k^2}\right]
\Biggr\}.
\label{growthrate_Q}
\ee
As previously mentioned, the first term, which contains the
differential $\left(\partial s/\partial\rho\right)_T<0$, is 
responsible for diffusive Silk damping.  The second term may
contribute to either damping or driving, depending upon the 
direction of propagation relative to the vertical.  
Upon inspection, eq. (\ref{growthrate}) indicates that 
for a given ${\bf k}$, only one of the compressible magnetosonic waves
is driven unstable, similar to the photon bubble instability.     

The relationship between the pressure and density 
perturbation of compressible MHD waves has allowed
us to calculate the damping and growth rates arising from the action of 
rapid thermal energy transfer in a stratified background, resulting
from coulomb diffusion along magnetic field lines.  At the same time, 
we realized that the work done upon a fluid element by the background 
heat flux is proportional to the overlap between the fluid displacement
$\bxi$ and the component of the perturbed heat flux that is $\perp\bhat$.
This clearly implicates $\delta{\bf Q}_{\perp}$ as the driving force 
responsible for the over-stability.  In what follows, we elaborate upon 
this point.     

\subsection{The Perpendicular Heat Flux, $\delta{\bf Q}_{\perp}$, as the 
Linear Driving Force and the Geometry of the Driving Mechanism}

Instead of examining the relationship between changes in pressure
and volume for a driven magnetosonic wave, we consider the 
overlap of the driving force with the fluid motion itself.  
To isolate the linear driving force, it is convenient to work with
the Lagrangian displacement $\bxi$.  The Euler equation reads 
\be
\frac{\partial^2\bxi}{\partial t^2}\simeq -\left({\bf k}\cdot{\bf v}_A
\right)^2\bxi-\left(c^2_i+v^2_A\right){\bf k}\left({\bf k}\cdot\bxi\right)
+{\bf v}_A\left({\bf k}\cdot{\bf v}_A\right)\left({\bf k}\cdot\bxi\right)
+{\bf k}\left({\bf k}\cdot{\bf v}_A\right)\left(\bxi\cdot{\bf v}_A\right)
-i\,{\bf k}\left(\frac{1}{\rho}\left(\frac{\partial P}{\partial T}
\right)_{\rho}\delta T -c^2_i\,\bxi\cdot\bnabla
{\rm ln}\rho\right)-i\,{\bf g}\left({\bf k}\cdot\bxi\right).
\ee  
On the right hand side, the first four terms are responsible for the 
standard magnetosonic restoring forces while the other terms on the 
right hand side are responsible for secular driving and damping.
Explicitly, 
\be
 \left({\bf k}\cdot{\bf v}_A
\right)^2\bxi+\left(c^2_i+v^2_A\right){\bf k}\left({\bf k}\cdot\bxi\right)
-{\bf v}_A\left({\bf k}\cdot{\bf v}_A\right)\left({\bf k}\cdot\bxi\right)
-{\bf k}\left({\bf k}\cdot{\bf v}_A\right)\left(\bxi\cdot{\bf v}_A\right)
=k^2v^2_{\rm ph}\bxi,
\ee
where $v^2_{\rm ph}=\omega^2/k^2$ is the magnetosonic phase velocity given
by eq. (\ref{dispersion}).  This allows us to write
\be
\left(-\omega^2+k^2v^2_{\rm ph}\right)\bxi\simeq 
-\frac{i}{\rho}\left(\frac{\partial P}{\partial T}\right)_{\rho}
\left[{\bf k}\,\delta T+\left({\bf k}\cdot\bxi\right)\bnabla T\right]
+i\,c^2_i\,\bxi\times\left({\bf k}\times\bnabla{\rm ln}\rho\right).
\ee
Note that the last term that is $\propto\bnabla{\rm ln}\rho$ is 
$\perp\bxi$ and therefore does not factor into the secular 
forcing of a given magnetosonic wave.  It follows that
\be
\left(-\omega^2+k^2v^2_{\rm ph}\right)\bxi\simeq \delta{\bf f}
\ee 
where $\delta{\bf f}$ is responsible for secular driving and 
damping and is given by 
\be
\delta{\bf f}=-\frac{i}{\rho}\left(\frac{\partial P}{\partial T}
\right)_{\rho}
\left[{\bf k}\,\delta T+\left({\bf k}\cdot\bxi\right)\bnabla T\right].
\label{simpleEuler}
\ee 
As expected from eq. (\ref{e: simplegrowth}), 
forcing occurs as a result of changes in temperature.  If we look at the 
projection of $\delta{\bf f}$ along the fluid displacement $\bxi$, we see
that
\be
\bxi\cdot\delta{\bf f}=\frac{1}{\rho}\frac{\Delta\rho}{\rho}\left(\frac{\partial
P}{\partial T}\right)_{\rho}\Delta T.
\label{connection}
\ee
The connection between $\delta{\bf f}$ and the Lagrangian pressure 
perturbation $\Delta P$ is clear and apparent.  The left hand side of 
eq. (\ref{connection}) is the work done upon a fluid element irrespective
of whether or not the motions involved are compressible or not.  For the
physics considered here, the driving force may be thought of as arising from 
changes in pressure due to changes in temperature, dictated by 
the flow of heat as prescribed by the first law of thermodynamics.  In the 
previous section we found it useful to divide the Lagrangian pressure
perturbation into a component solely responsible for diffusive Silk-like 
damping
and one that may potentially lead to over-stable driving.  Likewise, it is 
equally useful to decompose the secular driving force $\delta{\bf f}$ into 
its analogous portions.  With the help of eqs. (\ref{Tdamp}) 
and (\ref{Tdrive}) we have
\be
\delta{\bf f}_{\rm damp}\simeq -\frac{i}{\rho}\left(\frac{\partial P}
{\partial T}\right)_{\rho}{\bf k}\,\delta T^{\rm damp}=
\frac{1}{\rho}\left(\frac{\partial P}
{\partial T}\right)_{\rho}\frac{n\,\rho\,T\omega}{\left({\bf k}\cdot\bhat
\right)^2\chi}
\left(\frac{\partial s}{\partial\rho}\right)_T{\bf k}\,
\frac{\delta\rho}{\rho}
\label{fdamp}
\ee     
and
\be
\delta{\bf f}_{\rm drive}&\simeq&-\frac{i}{\rho}\left(\frac{\partial P}
{\partial T}\right)_{\rho}\Biggl\{{\bf k}\,\delta T^{\rm drive}+
\left({\bf k}\cdot\bxi\right)\bnabla T \Biggl\}\nonumber\\
&\simeq&-\frac{i}{\rho}\left(\frac{\partial P}
{\partial T}\right)_{\rho}\Biggl\{ -{\bf k}\left[
\frac{\left({\bf k}\cdot\bxi\right)\left(
\bhat\cdot\bnabla T\right)}{\left({\bf k}\cdot\bhat\right)}+
\frac{{\bf k}\cdot
\left(\bxi\cdot\left({\bf 1}-\bhat\bhat\right)\right)\left(\bhat\cdot
\bnabla T\right)}{\left({\bf k}\cdot\bhat\right)}
+\bxi\cdot\left({\bf 1}-\bhat\bhat\right)
\cdot\bnabla T\right] + \left({\bf k}\cdot\bxi\right)\bnabla T   \Biggr\}
\nonumber\\
&\simeq&\frac{i}{\rho}\left(\frac{\partial P}
{\partial T}\right)_{\rho}\Biggl\{ 2\,{\bf k}\left[
\frac{\left({\bf k}\cdot\bxi\right)\left(
\bhat\cdot\bnabla T\right)}{\left({\bf k}\cdot\bhat\right)}
-\left(\bhat\cdot\bxi\right)\left(\bhat\cdot\bnabla T \right) \right] 
+{\bf k}\,\bxi\cdot\bnabla T -\left({\bf k}\cdot\bxi\right)\bnabla T\Biggr\}.
\ee
The last two terms in the $\{\}$ are equal to $\bxi\times\left({\bf k}\times
\bnabla T\right)$, which is $\perp\,\bxi$ and cannot therefore, drive 
the fluctuations.  Now, the linear driving force 
may be written as
\be
\delta{\bf f}_{\rm drive}&\simeq& \frac{2i}{\rho}\left(\frac{\partial P}
{\partial T}\right)_{\rho}\frac{{\bf k}{\bf k}}{\left({\bf k}\cdot\bhat
\right)}\cdot\bxi\cdot\left({\bf 1}-\bhat\bhat\right)\left(\bhat\cdot
\bnabla T\right)= \frac{2}{\rho}\left(\frac{\partial P}
{\partial T}\right)_{\rho}\frac{{\bf k}{\bf k}}{\left({\bf k}\cdot\bhat
\right)^2}\cdot\delta\bhat\left(\bhat\cdot\bnabla T\right)=
-\frac{2}{\rho}\left(\frac{\partial P}
{\partial T}\right)_{\rho}\frac{{\bf k}{\bf k}\cdot{\bf\delta Q}_{\perp}}{\chi
\left({\bf k}\cdot\bhat\right)^2}.
\label{fdrive}
\ee
The form of $\delta{\bf f}_{\rm drive}$ in terms of the perturbed heat 
flux $\perp$ to $\bhat$, given by $\delta{\bf Q}_{\perp}=\delta{\bf Q}\cdot
\left({\bf 1}-\bhat\bhat\right)$, allows us to plainly interpret the 
driving resulting from anisotropic coulomb diffusion.  That is, 
$\delta{\bf f}_{\rm drive}$ results from the flux of energy that is 
$\perp$ to $\bhat$, but projected along the wave vector ${\bf k}$.

\subsection{Essence of the Driving Mechanism and a 
Comparison with Photon Bubbles}
\label{ss: essence}

The growth rate for photon-bubble driving can identically be found by 
using eq. (\ref{e: simplegrowth}), but with an Eulerian temperature 
perturbation $\delta T$ given by eq. (\ref{deltaTrad}) rather than 
eq. (\ref{deltaT}).  In a sense, the {\it thermodynamics} of compressible 
waves being driven by rapid thermal diffusion is similar in both 
cases.  However, the {\it mechanics} and geometry of 
the respective driving mechanisms draws a definite distinction 
between the two instabilities. 

Previously, we compared the temperature 
perturbation in the limit of rapid coulomb diffusion to that of 
rapid radiative diffusion in order to distinguish between the physics 
of photon bubbles and of the coloumb bubble 
instability whose growth rate is given by
eq. (\ref{growthrate}).  For comparison, the photon bubble 
growth rate reads  
\be
A_0=\frac{i\,{\tilde\omega}^2}{2\omega\left[2\omega^2-
k^2\left(c^2_i+v^2_A\right)\right]}\times\Biggl\{
\frac{1}{\rho\chi}\left(
\frac{\partial P}{\partial T}\right)_{\rho}\left[\rho\,n\,T\,\omega
\left(\frac{\partial s}{\partial\rho}\right)_T
+\frac{\left({\bf k}\cdot
{\bf v_A}\right)}{{\tilde\omega}^2}\left({\bf k}\times{\bf v_A}\right)\cdot
\left({\bf k}\times{\bf F}\right)\right] \Biggr\},
\label{PB_growthrate}
\ee
which is completely equivalent to the one-temperature MHD growth rates
given by eqs. (93) and (107) of BS03.  Note that the above expression 
is quite similar to the growthrate given by eq. (\ref{growthrate}). 
However, the difference between the driving mechanism of the 
photon bubble instability and the coulomb bubble instability 
is clear upon examination of the driving component of the 
temperature perturbation for photon bubbles
\be
\Delta T^{\rm drive}\simeq -\frac{\bxi\cdot\delta{\bf F}_{\perp}}
{\chi\delta\rho/\rho}\,\,\,\,\,\,{\rm PHOTON\,\,\,\, BUBBLES}.
\ee  
In this case, $\delta{\bf F}_{\perp}= \left({\bf 1}-{\bf k}{\bf
k}/k^2\right) \cdot\delta{\bf F}$ represents the component of
radiative (heat) flux perturbation that is $\perp$ to the wave vector
${\bf k}$.

\begin{figure}
\begin{center}
\includegraphics{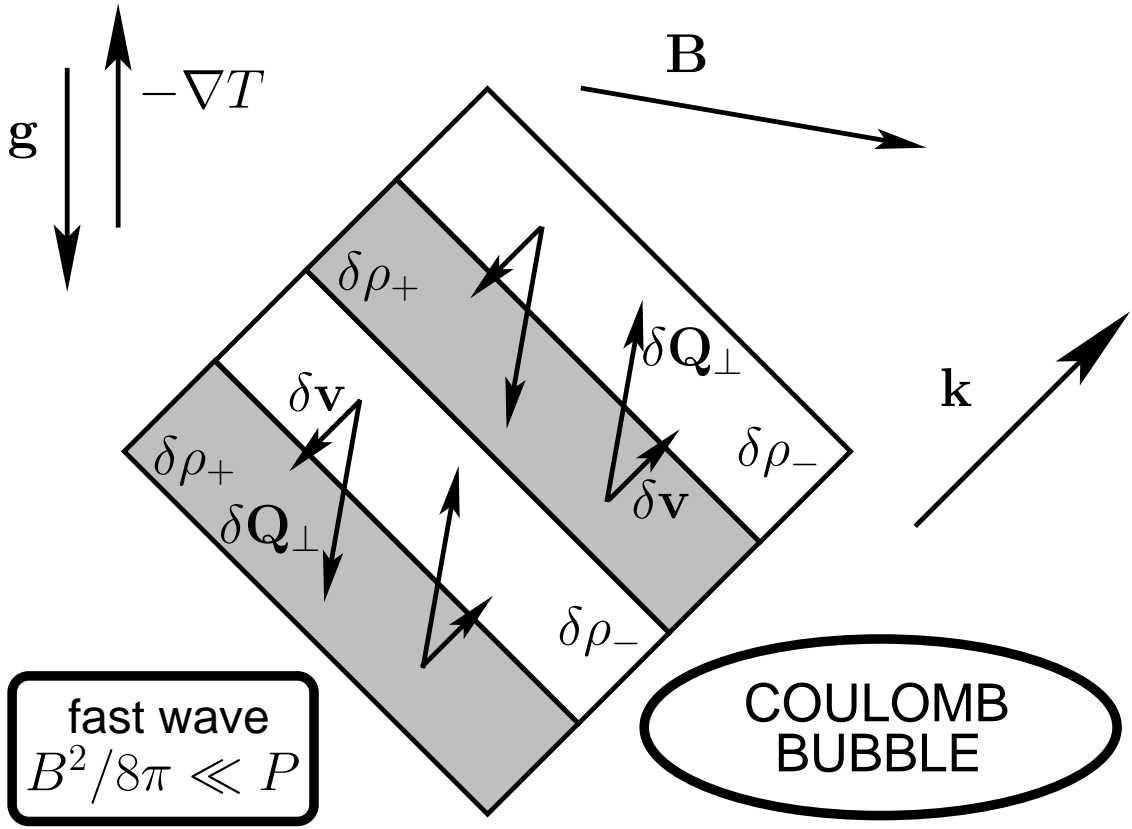}
\caption{Geometry of the coulomb bubble instability.  Here, the 
eigenfunction for a fast wave in a plasma with $B^2/8\pi/P\ll 1
$ -- nearly a standard hydrodynamic sound wave -- is depicted.  The velocity 
perturbation is perpendicular to surfaces of constant density and the 
perturbation of the perpendicular heat flux $\delta {\bf Q}_{\perp}$
lies along $\delta\hat{\bf b}$, the perturbation of field direction.
Secular driving occurs because the wave in question possesses a finite
projection along $\delta\hat{\bf b}$, which is $\perp$ to ${\bf B}$.
Thus, both the fast and slow waves are driven by the coulomb bubble mechanism 
whether or not the plasma is weakly magnetized or not.  The most 
strongly driven case corresponds to the example above i.e., a fast wave in a 
weakly magnetized plasma.}
\label{f:coulomb_pic}
\end{center}
\end{figure}

\begin{figure}
\begin{center}
\includegraphics{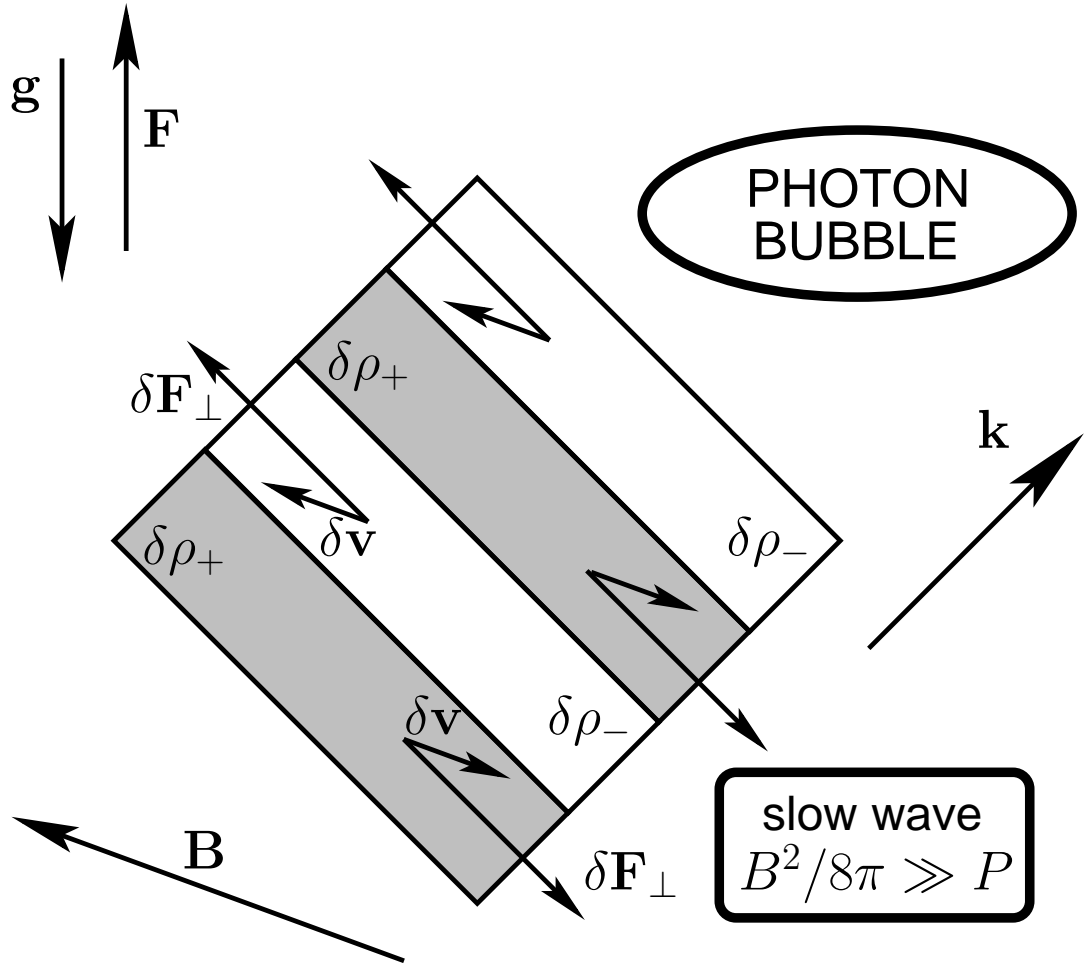}
\caption{Geometry of the photon bubble instability.  Both the 
fast and slow waves are driven by photon bubble driving.  
The most strongly driven case -- represented above -- is the slow
wave in a strongly magnetized plasma.}
\label{f:PB_pic}
\end{center}
\end{figure}

Figures \ref{f:coulomb_pic} and \ref{f:PB_pic} display the
geometry of the coulomb bubble instability and its relation to the
photon bubble instability.  Consider a fast magnetosonic
wave in a stratified plasma where $B^2/8\pi\ll P$ in the limit of
rapid anisotropic conduction.  That is, the wave can be roughly
thought of as a standard isothermal hydrodynamic sound wave with 
polarization vector $\hat{\bf \epsilon}\left(\omega\, ,
k\right)\parallel {\bf k}$ with an oscillation 
frequency $\omega\simeq k\,c_i$.  As
this acoustic disturbance propagates throughout the atmosphere, the
fluid motion along with the constraint of magnetic flux freezing
induces a magnetic field perturbation $\delta{\bf B}$.  In general,
$\delta {\bf B}$ possesses a component that is $\perp$ to the
equilibrium field ${\bf B}$, which then leads to a component of the
heat flux $\delta {\bf Q}_{\perp}$ that is $\perp{\bhat}$.  It is this 
component of the linear heat flux $\delta {\bf Q}$ that is responsible 
for the over-stable driving mechanism of the coulomb bubble instability.  
As long as the component of the velocity perturbation $\delta{\bf v}$
along ${\bf k}$ is endowed with a component that is $\perp$ to the
equilibrium field, then driving may ensue in the event that Silk damping 
is overcome.

Compare the geometry of the coulomb bubble mechanism described above
with that of the photon bubble instability.  Furthermore, consider
Table 2, which compares the growth rates of
both the coulomb bubble and photon bubble instability under various
equilibrium conditions.  Apparently, the coulomb bubble instability may
be thought of as the {\it mirror image} of the photon bubble
instability.  The prime mover for coulomb bubbles i.e., the physical
quantity responsible for the driving, is the component of the heat
flux perturbation $\delta{\bf Q}_{\perp}$ that is $\perp$ to the
equilibrium field ${\bf B}$.  For photon bubbles, driving originates
from the component of the radiative heat flux perturbation $\delta
{\bf F}_{\perp}$ that is $\perp$ to the wave vector ${\bf k}$.
Coulomb bubbles require that the component of the velocity perturbation
$\delta {\bf v}$ that is $\parallel$ to ${\bf k}$ possess some finite
projection with the perpendicular heat flux $\delta {\bf Q}_{\perp}$,
which is $\perp$ to ${\bf B}$.  For photon bubbles, the requirement is
that a component of the velocity perturbation $\delta {\bf v}$ that is
$\parallel$ to ${\bf B}$ possess some finite projection with the
perpendicular radiative heat flux $\delta {\bf F}_{\perp}$, which is
$\perp$ to ${\bf k}$.  The most strongly-driven coulomb bubble wave is
the {\it fast} wave in the limit where $p\gg B^2/8\pi$ (roughly a standard
hydrodynamic sound wave) such that the velocity perturbation 
$\delta{\bf v}$ is almost purely $\parallel$ to the wave vector 
${\bf k}$.  Finally, the most strongly-driven photon bubble wave is the 
{\it slow} wave in the limit where $p\ll B^2/8\pi$ such that the 
velocity perturbation $\delta{\bf v}$ is almost purely $\parallel$
to equilibrium field ${\bf B}$.

\begin{deluxetable}{ccccc}
%\rotate
\tabletypesize{\scriptsize}
\label{t: growthrates}
\tablecaption{Physical Properties and Conditions for Driving
for both Coulomb Bubbles 
and Photon Bubbles.\tablenotemark{a} \label{tbl-1}}
\tablewidth{0pt}
\tablehead{\colhead{Mode Branch} & \colhead{Diffusion Law} 
&\colhead{Plasma Beta} &\colhead{Instability Criterion} & 
\colhead{Asymptotic Growth Rate}}
\startdata 
FAST & ${\bf Q}=-\chi\bhat\bhat\cdot
\bnabla T$  &$p\gg B^2/8\pi$ &
$Q\gta p\,c_i$ & $\left(\frac{g}{c_i}\right)$ 
 \\
SLOW  & ${\bf Q}=-\chi\bhat\bhat\cdot
\bnabla T$   &$p\gg B^2/8\pi$  &
$Q\gta \left(\frac{v^2_A}{c^2_i}\right)\,p\, v_A $ &
$ \left(\frac{v_A}{c_i}\right)\left(\frac{g}{c_i}\right)$ 
 \\
FAST & ${\bf Q}=-\chi\bhat\bhat\cdot
\bnabla T$  &$p\ll B^2/8\pi$ &
$Q\gta p\,v_A$ &
$\left(\frac{c_i}{v_A}\right)\left(\frac{g}{c_i} \right)$ 
 \\
SLOW &${\bf Q}=-\chi\bhat\bhat\cdot
\bnabla T$  &$p\ll B^2/8\pi $ &
$Q\gta \left(\frac{v^2_A}{c^2_i}\right)\,p\,c_i $ &
$\left(\frac{c^2_i}{v^2_A}\right)\left(\frac{g}{c_i}\right)$
\\
FAST & ${\bf F}=-\chi\bnabla T$   &$p\gg B^2/8\pi$ &
$F\gta \left(\frac{c^2_i}{v^2_A}\right) p\,c_i$ & $\left(
\frac{v^2_A}{c^2_i}\right)\left(\frac{g}{c_i}\right)$ 
 \\
SLOW  & ${\bf F}=-\chi\bnabla T$ & $p\gg B^2/8\pi$  &
$F\gta \left(\frac{v^2_A}{c^2_i}\right)\,p\, v_A $ &
$ \left(\frac{v_A}{c_i}\right)\left(\frac{g}{c_i}\right)$ 
 \\
FAST & ${\bf F}=-\chi\bnabla T$ &$p\ll B^2/8\pi$ &
$F\gta p\,v_A$ &
$\left(\frac{c_i}{v_A}\right)\left(\frac{g}{c_i} \right)$ 
 \\
SLOW & ${\bf F}=-\chi\bnabla T$  &$p\ll B^2/8\pi $ &
$F\gta p\,c_i $ &$\left(\frac{g}{c_i}\right)$
 \enddata
\tablenotetext{a}{These relations are {\it approximate} and only 
accurate to the order unity level.  Terms that convey the 
geometric relationship between  ${\bf k}$, $\hat{\bf b}$, and the 
vertical gradient have been dropped.}
\end{deluxetable}

\section{An example: Interstellar Cosmic ray diffusion}
\label{s: examples}

We now understand the inner workings of the coulomb bubble 
driving mechanism and the equilibrium conditions 
in which such driving can operate.  Now, we put our analysis 
to use with the example of interstellar cosmic ray diffusion.

\subsection{Interstellar Cosmic Ray Diffusion and the 
``Two-Temperature'' Approximation}

Throughout the analysis above, we assert that the thermodynamics of 
the coulomb bubble instability is identical to that of the photon 
bubble instability.  Though this statement is valid, it is not 
entirely complete.   Blaes \& Socrates (2003) considered the 
possibility of photon bubble driving in two distinct thermodynamic
regimes.  If the micro-physical processes responsible for absorption 
and emission act quickly enough to maintain thermal equilibrium 
between the fluctuating radiation and gaseous particle field, then 
the thermodynamics of photon driving and damping are encapsulated
by a common temperature.  In other words, absorption 
and emission occur rapidly in comparison to the dynamical time
so that the radiation and gas temperature are locked to one another.
So far, our analysis of the coulomb bubble has taken place 
under this ``one-temperature'' approximation.  

If the particles endowed with a relatively high level of thermal 
mobility cannot come into close thermal contact with the 
bath of gaseous particles on a dynamical time, then the 
thermodynamics of radiative driving and damping is best described
by a ``two-temperature'' approximation.  An example of a system that
is susceptible to photon bubble driving in the two-temperature 
limit is a magnetized envelope that is optically thick and whose only
source of opacity is Thomson scattering.  

Interstellar cosmic ray diffusion is another example of where the 
transfer of a radiative energy is best described in a 
two-temperature approximation.  That is, the Galactic distribution 
of cosmic rays cannot come into local thermal equilibrium with the 
distribution of gaseous particles due to the absence of
absorption and emission processes between the two species.
Therefore, the thermodynamic state of the combined 
matter + cosmic ray fluid is constrained by the first law 
of thermodynamics for the matter gas given by 
eq. (\ref{ae: 1stlawgas}) in addition to eq. (\ref{ae: 1stlawCR}),
the cosmic ray energy equation.  

Though cosmic rays cannot directly exchange energy with
the matter gas, they certainly can exchange momentum via resonant 
scattering with magnetic irregularities on the Larmor scale. 
For $\sim 1-10$ GeV cosmic ray protons -- the energy range 
that is responsible for the majority of the Galactic 
cosmic ray pressure -- the Larmor radius $r_{_{L}}$
is given by
\be
r_{_L}\simeq 3\times 10^{12}\frac{\left(E_{_{\rm CR}}/{\rm GeV}\right)}
{\left(B/\mu{\rm G}\right)}\,\,{\rm cm},
\label{e: Larmor}
\ee
where $E_{_{\rm CR}}$ and $B$ is the cosmic ray energy and 
the large scale magnetic field strength, respectively.  
It follows that the length scale $r_{_L}$ of the resonant magnetic 
irregularities are much smaller than the inferred 
energy-weighted cosmic ray mean free path $\lambda_{_{\rm CR}}
\sim 1$ pc (Ginzburg et al. 1980; Strong \& Moskalenko 1998) 
and the scale height of the Galactic corona, which is
$\sim$ a few kpc. 

The cosmic ray pressure perturbation $\delta P_{_{\rm CR}}$ governs
the exchange of momentum between fluctuations in the cosmic ray
distribution and magnetofluid perturbations.  The linearized 
Euler with the inclusion of the cosmic ray pressure force
reads 
\be
-i\,\omega\rho\delta{\bf v}=-i\,{\bf k}\left[\delta P+\delta P_{_{\rm CR}}
\right]+{\bf g}\,\delta\rho+\frac{i}{4\pi}\left({\bf k}\times\delta
{\bf B}\right)\times {\bf B}=-i\,{\bf k}\left[\left(\frac{\partial
P}{\partial\rho}\right)_T\delta\rho+\left(\frac{\partial P}{\partial T}
\right)_{\rho}\delta T+\delta P_{_{\rm CR}}
\right]+{\bf g}\,\delta\rho+\frac{i}{4\pi}\left({\bf k}\times\delta
{\bf B}\right)\times {\bf B}.
\label{e: CR_Euler}
\ee
If Galactic cosmic rays were a ``normal'' radiation species
that could come into close thermal contact the matter through
absorption and emission processes, then local thermal equilibrium
between the radiation and matter distribution may occur and 
then both radiation and matter distributions may be characterized 
by a common temperature $T$.  Upon perturbation, both radiation 
and particle species share the same temperature perturbation $\delta T$
as long
as emission and absorption processes act quickly on dynamical 
timescales of interest (Blaes \& Socrates 2003).  For interstellar
cosmic ray diffusion, the absence of any absorption and emission 
opacity necessarily implies that cosmic ray pressure 
perturbation $\delta P_{_{\rm CR}}$ and gas temperature 
perturbation $\delta T$ are separate and decoupled.

Below, we discuss the 
possibility that the diffusion of interstellar cosmic rays 
in the Galactic corona leads to a ``two-temperature'' version 
of the coulomb bubble instability.    

\subsection{Pressure Perturbation, Stability Criteria
and Growth Rate}

The diffusion of interstellar cosmic rays in the galactic disk and
halo may lead to the two-temperature analogue of the photon bubble
instability.  In this case, cosmic rays, rather than photons, are
responsible for the diffusive radiative transfer of energy.  The 
action of cosmic
rays scattering off of resonant magnetic irregularities leads to an
``opacity'' responsible for mediating momentum exchange between the
radiation species and the gas.  In the photon case, the 
relevant interaction is governed by Thomson scattering of
photons with electrons.  Rather than the radiation field diffusing
along the gradient in the radiation pressure, cosmic rays diffuse
along the projection of the cosmic ray pressure gradient that 
coincides with the equilibrium magnetic field.  In what follows, 
we produce an abbreviated derivation of the growth rate
$A_0$ for the coulomb bubble instability resulting from 
cosmic ray diffusion in the galactic corona.

The perturbation of the cosmic ray conductivity is given by
\be
\frac{\delta\chi_{_{\rm CR}}}{\chi_{_{\rm CR}}}=
-\frac{\delta\rho}{\rho}-\frac{\partial{\rm ln}\kappa_{_{\rm CR}}}{
\partial{\rm ln}\rho}\frac{\delta\rho}{\rho}.
\label{e: chi_CR}
\ee
In Appendix \ref{a: cosmic rays}, we take note that $\partial 
{\rm ln}\kappa_{_{\rm CR}}/\partial{\rm ln}\rho\simeq -1$
for the Galactic halo.  The implication being that 
$\delta\chi_{_{\rm CR}}\simeq 0$.  However, we maintain the 
parameterization of $\delta\chi_{_{\rm CR}}$ given by 
eq. (\ref{e: chi_CR}) in order to separate the effects of a
``cosmic ray $\kappa$-mechanism'' and coulomb bubbles.\footnote{Note 
that the physical dimensions of the thermal 
conductivity $\chi$ and $\chi_{_{\rm CR}}$ are different.
Technically, $\chi_{_{\rm CR}}$ is a diffusivity rather 
than a conductivity.}  

The perturbation to the cosmic ray energy flux may be written as 
\be
\delta{\bf Q}_{_{\rm CR}}=-\chi_{_{\rm CR}}
\left[-\frac{\delta\rho}{\rho}\bhat\left(
\bhat\cdot\bnabla  P_{_{\rm CR}}\right)-\frac{\partial{\rm ln}
\kappa_{_{\rm CR}}}{\partial{\rm ln}\rho}\frac{\delta\rho}{\rho}
\bhat\left(\bhat\cdot\bnabla P_{_{\rm CR}}\right)  
+\delta\bhat
\left(\bhat\cdot\bnabla P_{_{\rm CR}}\right)+
\bhat\left(\delta\bhat\cdot\bnabla P_{_{\rm CR}}\right) 
+i\,\bhat\left({\bf k}\cdot\bhat\right)\delta P_{_{\rm CR}} \right],
\label{e: CR_fluxq}
\ee   
which then allows us to determine the cosmic ray pressure
perturbation $\delta P_{_{\rm CR}}$ in terms of $\delta\rho$
and $\delta\bhat$.  The linear cosmic ray energy equation reads
\be
-i\,\omega\delta P_{_{\rm CR}}+\frac{4}{3}i\,\omega P_{_{\rm CR}}
\frac{\delta\rho}{\rho}-i\,\omega\bxi\cdot\left(\bnabla P_{_{\rm CR}}
-\frac{4}{3}\frac{P_{_{\rm CR}}}{\rho}\bnabla\rho\right)=
-i\,{\bf k}\cdot\delta{\bf Q}_{_{\rm CR}}.
\label{e: CR_energy}
\ee  
The second term on the left hand side in above expression 
leads to diffusive Silk damping, while the third term 
is responsible for Brunt-Vaisalla oscillations.  In 
\S\ref{ss: deltaT} we took the high-$k$ limit of rapid diffusion 
of the first law of thermodynamics 
in order to isolate the temperature perturbation $\delta T$
in terms of the density and magnetic field unit vector perturbation
$\delta \rho$ and $\delta\bhat$, respectively.  In order
to study coulomb bubbles in the ``two-temperature'' approximation 
appropriate for cosmic ray diffusion, we likewise 
isolate the cosmic ray pressure perturbation 
$\delta P_{_{\rm CR}}$ in terms of $\delta\rho$ and $\delta\bhat$
in the limit of rapid diffusion.  By combining 
eqs. (\ref{e: CR_fluxq}) and (\ref{e: CR_energy}), the cosmic 
ray pressure perturbation becomes
\be
\delta P_{_{\rm CR}} & \simeq & \frac{4}{3}\frac{\omega}{\omega
_{\rm CR, diff}}
P_{_{\rm CR}}\frac{\delta\rho}{\rho}-\frac{\omega}
{\omega_{\rm CR,diff}}\bxi\cdot\left(\bnabla P_{_{\rm CR}}
-\frac{4}{3}\frac{P_{_{\rm CR}}}{\rho}\bnabla\rho \right)\nonumber
\\
&-&\frac{\chi_{_{\rm CR}}}{\omega_{\rm CR,diff}}\left[
-\frac{\delta\rho}{\rho}\left(1+\frac{\partial\,{\rm ln}\kappa_{_{\rm CR}}}
{\partial{\rm ln}\rho} \right)\left({\bf k}\cdot\bhat\right)
\left(\bhat\cdot\bnabla P_{_{\rm CR}}\right)
+\left({\bf k}\cdot
\delta\bhat\right)\left(\bhat\cdot\bnabla P_{_{\rm CR}} \right)
+\left({\bf k}\cdot\bhat\right)\left(\delta\bhat\cdot\bnabla
P_{_{\rm CR}}\right)\right],
\ee
where $\omega_{\rm CR,diff}=\omega+i\chi_{_{\rm CR}}\left({\bf k}
\cdot\bhat\right)^2$ is the characteristic cosmic ray diffusion 
frequency.  In the limit of rapid diffusion, we take
\be
|\omega_{\rm CR, diff}|\rightarrow \,\chi_{_{\rm CR}}\left(
{\bf k}\cdot\bhat\right)^2\gg \omega. 
\ee 
Above, we take advantage of the fact that to leading order, the 
frequency of an acoustic oscillation is $\propto k$.  
With this, we have
\be
\delta P_{_{\rm CR}}& \simeq -&\frac{4i}{3}\frac{\omega\,
P_{_{\rm CR}}}{\chi_{_{\rm CR}}\left({\bf k}\cdot\bhat\right)^2}
\frac{\delta\rho}{\rho}+\frac{i}{\left({\bf k}\cdot\bhat\right)^2}
\left[-\frac{\delta\rho}{\rho}\left(1+
\frac{\partial\,{\rm ln}\kappa_{_{\rm CR}}}{\partial{\rm ln}\rho}\right)
\left({\bf k}\cdot\bhat\right)\left(
\bhat\cdot\bnabla P_{_{\rm CR}}\right)+\left({\bf k}\cdot
\delta\bhat\right)\left(\bhat\cdot\bnabla P_{_{\rm CR}} \right)
+\left({\bf k}\cdot\bhat\right)\left(\delta\bhat\cdot\bnabla
P_{_{\rm CR}}\right)\right].
\label{e: deltaPcr}
\ee
The similarity is evident between $\delta P_{_{\rm CR}}$ above and the
two-temperature radiation pressure perturbation $\delta P_{\rm rad}$ 
given by Blaes \& Socrates (2003)
\be
\delta P_{\rm rad} =\frac{1}{3}\,\delta E=
-\frac{i}{k^2}\left[\frac{4E}{3}\frac{\omega}{\chi_R}
\frac{\delta\rho}{\rho}
+\frac{\delta\rho}{\rho}\left({\bf k}\cdot\bnabla P_{\rm rad}\right)
\left(1+\frac{\partial\,{\rm ln\kappa_F}}
{\partial\,{\rm ln}\rho}\right)\right],\,\,\,\,\,\,\,\,
{\rm 2-TEMP\,\,\,\, RADIATIVE\,\,\,\, DIFFUSION}
\label{e: deltaPrad}
\ee
where $\kappa_F$ is the flux mean opacity as defined in 
Blaes \& Socrates (2003).  Note that the Silk damping term
of eq. (\ref{e: deltaPcr}) is smaller than the one found 
in eq. (\ref{e: deltaPrad}) by a factor of three, which results
from our choice of parameterization of the cosmic ray diffusivity 
$\chi_{_{\rm CR}}$.     
We are now in the position to calculate the growth rate and 
stability criteria for the cosmic ray coulomb bubble instability.
Following the analysis that led to eq. (\ref{e: simplegrowth}), 
the ratio of the asymptotic growth rate to acoustic 
oscillation frequency becomes
\be
\frac{A_0}{\omega}\simeq\frac{1}{2}\frac{\frac{\Delta\rho}{\rho}
\Delta P_{_{\rm CR}}}{\rho\,\delta{\bf v}\cdot\delta{\bf v}}.
\label{e: CR_growth1}
\ee 
Note that in the two-temperature case the velocity eigenvector
oscillates at magnetoacoustic frequencies that are the solution to 
eq. (\ref{e: magnetosonic}).  The quantitative difference
is that the isothermal gas sound speed $c_i$ is replaced with
the adiabatic gas sound speed $c_s$, since the gas neither
generates nor loses any heat itself, the gas temperature 
perturbation then contributes to the acoustic response.  
In terms of its individual 
constituents, the growth rate may be written as
\be
A_0\simeq \frac{{\tilde\omega}^2k^2}{2\omega\left[2\omega^2
-k^2\left(c^2_s+v^2_A \right)\right]}\times
\left[\frac{i}{\rho\chi_{_{\rm CR}}\left({\bf k}\cdot
\bhat\right)^2}\right]\Biggl\{-\frac{4}{3}
\omega P_{_{\rm CR}}+\frac{2i\,{\tilde\omega}^2k^2}{\omega^2}
\frac{\bxi\cdot\delta{\bf Q}_{_{\rm CR,}\perp}}{\left(
\delta\rho /\rho\right)^2} 
+\frac{\partial\,{\rm ln}\kappa_{_{\rm CR}}}{\partial{\rm ln}
\rho}\,
\left({\bf k}\cdot{\bf Q}_{_{\rm CR}}\right)\Biggl\}.
\label{e: CR_growth2}
\ee    
The form of $A_0$ not only serves as an asymptotic growth rate,
but as a stability criteria as well.
The first term in the 
$\{\}$ leads to Silk damping, the second term results in 
the coulomb bubble instability and the last term is responsible
for a $\kappa_{_{\rm CR}}-$mechanism.  The quantity $\delta
{\bf Q}_{_{\rm CR},\perp}=\left({\bf 1}-\bhat\bhat\right)\cdot
\delta{\bf Q}_{_{\rm CR}}$ is the cosmic ray flux perturbation 
that is $\perp$ to the equilibrium field ${\bf B}$ (and equilibrium 
cosmic ray flux ${\bf Q}_{_{\rm CR}}$). 
It follows that the geometry and mechanics of the two-temperature cosmic 
ray coulomb bubble instability is identical to the 
one-temperature coulomb bubble instability.  

In terms of the wave vector and specified equilibrium
parameters, the asymptotic growth rate becomes
\be
A_0\simeq  \frac{{\tilde\omega}^2k^2}{2\omega\left[2\omega^2
-k^2\left(c^2_s+v^2_A \right)\right]}\times
\left[\frac{i}{\rho\chi_{_{\rm CR}}\left({\bf k}\cdot
\bhat\right)^2}\right]\Biggl\{-\frac{4}{3}
\omega P_{_{\rm CR}}+\frac{2\omega^2}{{\tilde\omega}^2}
\left({\bf k}\cdot{\bf Q}_{_{\rm CR}}\right)\left[1-
\frac{\left({\bf k}\cdot\bhat\right)^2}{k^2}\right]
+\frac{\partial\,{\rm ln}\kappa_{_{\rm CR}}}{\partial{\rm ln}
\rho}
\left({\bf k}\cdot{\bf Q}_{_{\rm CR}}\right)\Biggl\}.
\label{e: CR_growth3}
\ee

\subsection{Marginal Stability of the Galactic Corona}
     
As stated in Appendix \ref{a: cosmic rays}, Galactic cosmic rays
almost uniformly fill a halo (or corona) with a characteristic
thickness $\sim$ a few kpc.  Over these scales, the coronal gas is hot
($T\sim 10^6$ K) and tenuous $n\sim 10^{-3}\,{\rm cm^{-3}}\equiv n_3$
and the corresponding adiabatic gas sound speed $c_s\sim 10^7
T^{1/2}_6{\rm cm/s}$, where $T_6$ is temperature in units of $10^6$ K.
For our stability analysis, we take the halo equilibrium field to be
uniform with a characteristic value of $B\sim 1\mu{\rm G}\equiv
B_{{\mu{\rm G}}}$, corresponding to an Alfv\'en velocity $v_A\lesssim
10^7 B_{{\mu{\rm G}}}\,n^{-1/2}_3 {\rm cm/s}$, somewhat less than the
adiabatic sound speed $c_s$.  The average local pressure for gas and
cosmic ray protons in the interstellar medium is roughly $\sim 1{\rm
eV/cm^3} \sim 10^{-12}\,{\rm erg/cm^3}$.  Due to the large $\sim$ a few kpc
scale height for the cosmic rays, this value of pressure also
corresponds to the coronal value of the cosmic ray pressure $P_{_{\rm
CR}}\sim 1{\rm eV/cm^3}$.  Compare this with the value of the coronal
gas pressure $P\sim 10^{-1}\,n_3 \,T_6\, {\rm eV/cm^3}$.  Thus, the
ratio of cosmic ray to gas pressure is $P_{_{\rm CR}}/P\sim 10$ in the
kpc-scale galactic corona, a somewhat surprising value.  In short, 
the Galactic corona is a cosmic ray pressure supported 
atmosphere.  
  
In the diffusion approximation, the cosmic ray flux $Q_{_{\rm CR}}$
for a plane-parallel geometry satisfies the relation
\be
\frac{\tau_{_{\rm CR}}}{c}Q_{_{\rm CR}}\sim P_{_{\rm CR}}.
\ee
For a disk-like geometry the corresponding cosmic ray luminosity 
$L_{_{\rm CR}}$ is roughly
\be
L_{_{\rm CR}}\sim 5\times 10^{40}P_{1}\,\tau^{-1}_{3.5}
\,R^2_{10}\,{\rm erg\,s^{-1}}
\ee
where $\tau_{3.5}$, $P_{1}$ and $R_{10}$ is the cosmic ray optical
depth in units of $10^{3.5}$, halo cosmic ray pressure in units of
$1\,{\rm eV/cm^3}$ and characteristic cylindrical emitting radius in
units of 10 kpc, respectively.  We now have the necessary ingredients
to determine whether or not the kpc-scale galactic corona is
over-stable to either the coulomb bubble instability or 
the $\kappa$-mechanism.  When calculating the growth or damping 
rates resulting from cosmic ray diffusion, we assume that the 
equilibrium field is relatively uniform over a wavelength 
and that it does not contribute to hydrostatic balance
as well.  

In \S\ref{ss: essence} we made note of the fact that the 
coulomb bubble instability is most strongly driven for
waves that closely resemble simple propagating hydrodynamic 
sound waves i.e., the fast wave in the limit where $c^2_s\gg v^2_A$.   
In this case, the {\it instability} condition for cosmic ray 
coulomb bubble driving from eq. (\ref{e: CR_growth3}) is 
approximately
\be
Q_{_{\rm CR}} & > & P_{_{\rm CR}}\, c_s
\longrightarrow \frac{c}{\tau_{_{\rm CR}}}>c_s.
\ee
In the above expression we ignore geometrical and other
factors of order unity. Interestingly, the quantity 
$c/\tau_{_{\rm CR}}$, which represents the diffusive drift velocity 
$v_{\rm D}$ over the coronal scale height, is close in value
to the adiabatic gas sound speed i.e., $v_D\sim 
\tau^{-1}_{3.5}\,10^7{\rm cm/s}\sim c_s\sim 
10^7T^{1/2}_6{\rm cm/s}$.  Altogether {\it the condition for 
instability is satisfied as long as} $T^{1/2}_6\tau_{3.5}<1$.
The growth rate $\Gamma_{_{\rm FAST}}$ 
for the fast wave coulomb bubble instability in 
the $c^2_s\gg v^2_A$ limit is given by
\be
\Gamma_{_{\rm FAST}}\sim \frac{g}{c_s}\sim
\frac{c_{_{\rm CR}}}{c_s}\frac{g}{c_{_{\rm CR}}}\sim
\sqrt{\frac{P_{_{\rm CR}}}{P}}\frac{V_{\rm circ}}{H_c}
\sim 3\times 10^{-7}\,\sqrt{\frac{P_{_{\rm CR}}/P}{10}}
\,V_{300}\,H^{-1}_3\,{\rm yrs^{-1}},
\ee
where $V_{300}$ is the circular velocity $V_{\rm circ}$
in units of $300\,{\rm km/s}$ and $H_3$ is the coronal
scale height in units of 3 kpc.  The growth rate 
$\Gamma_{_{\rm FAST}}$ is dynamical and indicates that the 
fast wave coulomb bubble instability can transform a low amplitude
propagating fast wave, with phase velocity equal to $c_s$, into 
a large amplitude sonic disturbance over a relatively few wave 
crossing times.     
 
\subsection{Coulomb Bubbles vs. the $\kappa$-Mechanism}

The approximate stability criteria and growth rate 
for a $\kappa_{_{\rm CR}}-$
mechanism is identical to that of the cosmic ray coulomb 
bubble instability in the corona of the Galaxy.  Fast waves 
in the weakly magnetized limit, with $\omega\simeq k\,c_s$, 
are the strongest growing excitations for both cases.  However, 
eq. (\ref{e: CR_growth3}) informs us that upward propagating
fast waves are driven by the coulomb bubble instability while
downward propagating fast waves are driven by the 
$\kappa_{_{\rm CR}}-$mechanism, for $\partial{\rm ln}
\kappa_{_{\rm CR}}/\partial{\rm ln}\rho\simeq -1$.

At the base of the corona, the Galactic disk is highly inhomogeneous
and turbulent.  The turbulence in the interstellar medium primarily
results from explosive phenomena -- expanding HII regions, line-driven
stellar winds and core-collapse SNe -- that originates, in some way,
from massive stars.  There are $\sim 10^4$ O stars in the Milky Way
within a galactocentric radius of $\sim 10$ kpc.  Therefore, the
characteristic separation between luminous massive stars $\sim 100$ pc
is roughly the scale $l_T$ at which the interstellar turbulence is
stirred.\footnote{This approximation for $l_T$ is valid as 
long as the lifetime of massive stars is short in comparison to  
the dynamical timescales of interest.}  

With respect to the Galactic corona, the dense multiphase disk serves
as a source of acoustic radiation that can then be amplified in the
corona by either the cosmic ray coulomb bubble instability or the
$\kappa_{_{\rm CR}}-$mechanism.  The direction of propagation
for the seed $k^{-1}\sim l_T\sim 100$ pc-scale acoustic fluctuations
is primarily upwards with respect to gravity in the Galactic
corona since the dense disk lies at its base.  It follows
that the cosmic ray coulomb bubbles are the the relevant 
mechanism that amplifies propagating acoustic
perturbations arising from the Galactic disk rather than a
$\kappa_{\rm _{CR}}-$mechanism.

\section{Summary}
\label{s: summary}

We have identified a new acoustic over-stability, which we refer to as
the ``coulomb bubble instability,'' that is driven by the rapid
diffusion of energy along magnetic field lines.  From a linear
instability analysis, we calculate the condition for over-stability
and the growth rate of the coulomb bubble instability.  Driving occurs
when the ratio of heat flux to entropy is relatively large for the
equilibrium, implying that the heat flux is the physical quantity that
provides the free energy required for the over-stability.  The coulomb
bubble instability may be thought of as a standard magnetoacoustic
wave that is driven by $\delta {\bf Q}_{\perp}=\left({\bf 1}-
\bhat\bhat\right)\cdot\delta{\bf Q}$, the linear perturbation of the
coulomb heat flux that is $\perp$ to the direction of the equilibrium
field $\bhat ={\bf B}/B$ (and thus ${\bf
Q}=-\chi\bhat\bhat\cdot\bnabla T$ as well).  The growth rate of the
coulomb instability is dynamical for the fast wave in the $B^2/8\pi\ll
P$ limit -- roughly a standard hydrodynamic sound wave -- such that
the growth time is roughly the sound crossing time over a gas pressure
scale height.

The properties of the coulomb bubble instability are, at the same
time, strikingly similar and starkly different from the photon bubble
instability.  The thermodynamic properties of both instabilities are
identical in that the rapid diffusion of thermal energy in a
stratified flow is needed in order for driving to occur.  The work
done upon the fluid by the driving force for both instabilities
is proportional to the product of the Lagrangian density and
temperature perturbation $\Delta\rho\,\Delta T$.  Furthermore, and
like the photon bubble instability, both the fast and the slow wave
are susceptible to over-stability from the coulomb bubble mechanism
irrespective of whether or not the background flow is strongly
magnetized.

Work is done upon the wave by the coulomb bubble linear driving force
provided by $\delta{\bf Q}_{\perp}$, the component of the heat flux
perturbation that is $\perp\bhat$.  The fast and slow magnetosonic
waves are polarized along both the wavevector ${\bf k}$ due to
compression, and along the direction of the magnetic field ${\bf b}$,
which results from magnetic tension.  In general, $\delta{\bf
Q}_{\perp}$ has some finite projection that is coincident with the
wavevector ${\bf k}$ -- as long as ${\bf k}\nparallel\bhat$ -- which
allows driving to occur, since all compressible fluctuations 
exhibit motion that is in part $\parallel{\bf k}$.  On the other hand,
the photon bubble linear driving force originates from $\delta{\bf
F}_{\perp}=\left({\bf 1}- {\bf k}{\bf k}/k^2\right)\cdot\delta{\bf
F}$, the component of the radiative heat flux perturbation that is
$\perp{\bf k}$.  Since the magnetic tension force partially polarizes
magnetosonic waves along $\bhat$, the radiative driving force due to
$\delta{\bf F}_{\perp}$ is coincident with the motion of the fluid,
which then allows the radiation field to perform work upon the fluid.
Clearly, in describing the geometry of the driving mechanisms for the
coulomb bubble and photon bubble instability, the vectors ${\bf k}$
and $\bhat$ are interchanged with one another in every possible way.
From this, we conclude that the coulomb bubble instability is the {\it
mirror opposite} of the photon bubble instability.

Both the fast and slow magnetosonic waves are driven over-stable by
the coulomb bubble mechanism.  The fast mode in the weakly magnetized
limit is the most strongly driven case, with a growth rate $\sim
g/c_g$, where $c_g$ is the gas sound speed.\footnote{In the
one-temperature limit, the appropriate gas sound speed $c_g$ is given
by the isothermal sound speed $c_i$, whereas is the two-temperature
limit the relevant gas sound speed $c_g$ is given by the adiabatic
sound speed $c_s$.}  If the equilibria in question resembles a stellar
envelope, then both the coulomb bubble instability and Balbus'
magnetothermal instability possess comparable growth rates.  However,
if the particles responsible for the transfer of thermal energy along
field lines (e.g. cosmic rays) are not the particles
responsible for providing the acoustic response (the gas), then the
growth rate of the coulomb bubble instability is larger than the
magnetothermal instability by a factor that is $\sim c_{_{\rm CR}}/c_i
\sim\sqrt{P_{_{\rm CR}}/P}$, for the case of cosmic ray diffusion in
the Galactic halo.

The coulomb bubble instability may thrive in a wide variety of
astrophysical environments.  For example, it is possible that coulomb
bubbles play an important role in the modifying the thermal transport
properties of the envelopes of magnetized neutron stars, central
cooling flows of galaxy clusters and interstellar cosmic ray
diffusion.  In the latter case, we show that the kpc-scale
galactic corona is a setting in which a cosmic ray coulomb 
bubble instability possibly operates.  For commonly accepted 
values of cosmic ray and gas pressure, cosmic ray 
luminosity and coronal gas temperature, the Galactic
corona is in a state of marginal stability with respect
to a cosmic ray coulomb bubble instability.  The implication being 
that cosmic ray coulomb bubbles self-regulates the interstellar 
cosmic ray pressure on Galactic scales.     

\subsection{What Is To Be Done?}

Our local linear analysis is only the first step in realizing whether
or not the coulomb bubble instability has practical applications.  At
this simple exploratory level, we only examine local propagating WKB
waves at linear order.  The next step is to perform a global linear
analysis, where a linear mode is really the combination of an upward
and downward propagating wave, whose properties are further augmented
by the specified boundary conditions.  Of course, linear theory can
only tell us whether an instability exists.  A MHD computer algorithm
that incorporates the effects of rapid anisotropic diffusion in a
stratified atmosphere is most likely the correct tool in fully
studying the non-linear outcome of the coulomb bubble instability.

\acknowledgements{AS acknowledges support of a Hubble Fellowship
administered by the Space Telescope Institute and to the Institute for
Theoretical Physics at the University of California, Santa Barbara for
its hospitality as a portion of this work was completed there.  IP
is supported by a Department of Energy Computational Science
Graduate Fellowship.}

\appendix

\section{An envelope partially supported by Cosmic Rays}
\label{a: cosmic rays}

In order to describe the physics of cosmic ray diffusion 
in a stratified atmosphere, we amend the conservation laws and 
equilibrium conditions of \S\ref{s: basic}.
Conservation of momentum now reads
\be
\rho\left(\frac{\partial {\bf v}}{\partial t} +{\bf v}\cdot\bnabla
{\bf v}\right)=-\bnabla\left(P+P_{_{\rm CR}}\right)
+\rho\,{\bf g}+\frac{1}{4\pi}\left(\bnabla 
\times {\bf B}\right)\times{\bf B},
\label{ae: momentum}
\ee
where $P_{_{\rm CR}}$ is the cosmic ray pressure -- dominated 
by $\sim 1-10$ GeV cosmic ray protons in the interstellar 
medium.  The first law of thermodynamics for the gas 
\be
\rho T\left(\frac{\partial s}{\partial t} +{\bf v}\cdot\bnabla s\right)
=0
\ee
now implies that fluid-matter perturbations are adiabatic.  
Evolution of the cosmic ray pressure $P_{_{\rm CR}}$ is 
further constrained by a cosmic ray-energy equation
\be
\frac{\partial P_{_{\rm CR}}}{\partial t}+{\bf v}\cdot\bnabla
P_{_{\rm CR}}+\frac{4}{3}P_{_{\rm CR}}\bnabla\cdot{\bf v}=
-\bnabla\cdot{\bf Q}_{_{\rm CR}},
\label{ae: 1stlawgas}
\ee
where we assume that adiabatic index of the cosmic rays
is that of a relativistic ideal gas i.e., $\gamma_{_{\rm CR}}
=4/3$.  The cosmic ray energy flux ${\bf Q}_{_{\rm CR}}$ is given by
\be
{\bf Q}_{_{\rm CR}}=-\chi_{_{\rm CR}}\bhat\bhat\cdot\bnabla 
P_{_{\rm CR}},
\label{ae: 1stlawCR}
\ee  
which is similar in form to the coulombic heat flux given by eq. 
(\ref{e: heat flux}).  

The cosmic ray conductivity $\chi_{_{\rm CR}}$
deserves discussion as its prescription for the Galaxy differs
from the prescription for thermal conductivity $\chi$ described
in \S\ref{ss: heat flux}.  Measurements of the cosmic ray transport
in the Milky Way indicate a cosmic ray mean free path 
$\lambda_{_{\rm CR}}\sim 1$pc.  Interestingly, models of 
Galactic cosmic ray transport are consistent with 
$\lambda_{_{\rm CR}}$ roughly being a constant over a ``halo'' 
scale height of $\sim$ a few kpc.  It follows that the cosmic 
ray ``opacity'' $\kappa_{_{\rm CR}}$ must vary inversely
with density i.e., $\kappa_{_{\rm CR}}\propto\rho^{-1}$.  The 
immediate suggestion is that a ``cosmic ray $\kappa$-mechanism'' 
may operate in the galactic halo since $\partial{\rm ln}
\kappa_{_{\rm CR}}/\partial{\rm ln}\rho\neq 0$ and is 
given by
\be
\frac{\partial\, {\rm ln}\kappa_{_{\rm CR}}}{\partial\,
{\rm ln}\rho}\simeq -1.
\ee   
Finally, the only other basic assumption that differs
from those given in \S\ref{s: basic} involves
hydrostatic balance
\be
-\frac{1}{\rho}\bnabla\left(P+P_{_{\rm CR}}\right)={\bf g}.
\ee
That is, partial support of the atmosphere is given by 
an equilibrium cosmic ray pressure gradient.  

\end{document}